\documentclass[twocolumn]{aastex62}
\usepackage[single=true]{acro}
\usepackage{xspace,bm,upgreek,amsmath,nicefrac}
\graphicspath{{./}{figures/}}

\received{}
\revised{}
\accepted{}
\submitjournal{ApJ}

\shorttitle{High-Energy Particles in the SS433 Jets}
\shortauthors{Sudoh et al.}

\DeclareAcronym{vhe}{
  short = VHE ,
  short-plural-form = \aciffirst{VHE, $>100$GeV}{VHE} ,
  long  = very-high-energy ,
  class = astro ,
  first-style = default 
}

\DeclareAcronym{he}{
  short = HE ,
  long  = high-energy ,
  long-plural-form = high energies ,
  class = hydro ,
  first-style = default
}

\DeclareAcronym{ic}{
  short = IC ,
  long  = inverse Compton ,
  class = process ,
  first-style = default
}

\DeclareAcronym{fer}{
  short = {\it Fermi}/LAT ,
  long  = {\it Fermi}/Large Area Telescope ,
  class = object ,
  first-style = default
}

\newcommand{\vk}{\ensuremath{\xi}}
\newcommand{\be}{\begin{equation}}
\newcommand{\ee}{\end{equation}}
\newcommand{\fer}{\ac{fer}\xspace}

\newcommand{\sub}[1]{_{\rm #1}}
\newcommand{\inj}{_{\rm inj}}
\newcommand{\jet}{_{\rm jet}}
\newcommand{\bohm}{_{\rm Bohm}}
\newcommand{\acc}{_{\rm acc}}
\newcommand{\syn}{_{\rm syn}}
\newcommand{\ic}{_{\rm ic}}
\newcommand{\ad}{_{\rm ad}}
\newcommand{\Ge}{_L}
\newcommand{\Bf}{_B}

\begin{document}

\title{Multi-wavelength Emission from Galactic Jets: the Case of the Microquasar SS433}

\correspondingauthor{Takahiro Sudoh}
\email{sudoh@astron.s.u-tokyo.ac.jp}

\author[0000-0002-6884-1733]{Takahiro Sudoh}
\affil{Department of Astronomy, University of Tokyo, Hongo, Tokyo 113-0033, Japan}

\author[0000-0002-7272-1136]{Yoshiyuki Inoue}
\affil{Interdisciplinary Theoretical \& Mathematical Science Program (iTHEMS), RIKEN, 2-1 Hirosawa, Saitama 351-0198, Japan}
\affil{Kavli Institute for the Physics and Mathematics of the Universe (WPI), UTIAS, The University of Tokyo, Kashiwa, Chiba 277-8583, Japan}

\author[0000-0002-7576-7869]{Dmitry Khangulyan}
\affil{Department of Physics, Rikkyo University, Nishi-Ikebukuro 3-34-1, Toshima-ku, Tokyo 171-8501, Japan}

\begin{abstract}
SS433 is a Galactic microquasar with powerful jets, where very-high-energy particles are produced. We study particle acceleration in the jets of SS433 in the light of the recent multi-wavelength data from radio to TeV gamma ray. We first present a general framework for the particle acceleration, cooling, and transport in relativistic jets. We then apply this to two X-ray knots in the jets of SS433, focusing on leptonic emission. Our detailed treatment of particle transport and evolution produces substantially different predictions from previous papers. For both regions, our model can account for the multi-wavelength data except for the GeV data. This suggests that GeV emission is mostly from different regions and/or mechanisms. We find that the acceleration process should be efficient, which could be realized by diffusive shock acceleration close to the Bohm limit. Provided that protons are accelerated at the same efficiency as electrons, our results imply that SS433 is a PeVatron, i.e., can accelerate protons beyond a PeV. Future hard X-ray and MeV gamma-ray observations can critically test our models by detecting the spectral turnover or cutoff.
\end{abstract}

\keywords{acceleration of particles, astroparticle physics, black hole physics, gamma rays: general}

\section{Introduction} \label{sec:intro}
\acresetall
Microquasar SS433 is a binary system consisting of a compact object (most likely a black hole) and a supergiant star \citep[e.g.,][]{Hillwig04,Hillwig08,Kubota10,Cherepashchuk19}, which is believed to feed a super-critical accretion disk \citep[e.g.,][]{Begelman06,Medvedev10,Cherepashchuk13}. Thanks to its relative proximity \citep[$5.5$~kpc,][]{Blundell04,Lockman07} and a number of unique features, this object has been intensively studied for decades, though many aspects remain mysterious \cite[see][for a review]{Fabrika04}. A particularly striking feature is a pair of jets, which are launched almost perpendicular to the line of sight and show periodic precession and nodding motion  \citep{Abell79,Fabian79}. They are mildly relativistic ($v=0.26c$, where $c$ is the speed of light) and have plenty of power ($\sim 10^{39}$~erg~s$^{-1}$) \citep[e.g.,][]{Marshall02,Brinkmann05}. The jets interact with the surrounding nebula W50, believed to be a supernova remnant \citep[e.g.,][]{Dubner98,Green04}.  

\begin{figure*}[t]
\resizebox{18cm}{!}{\includegraphics{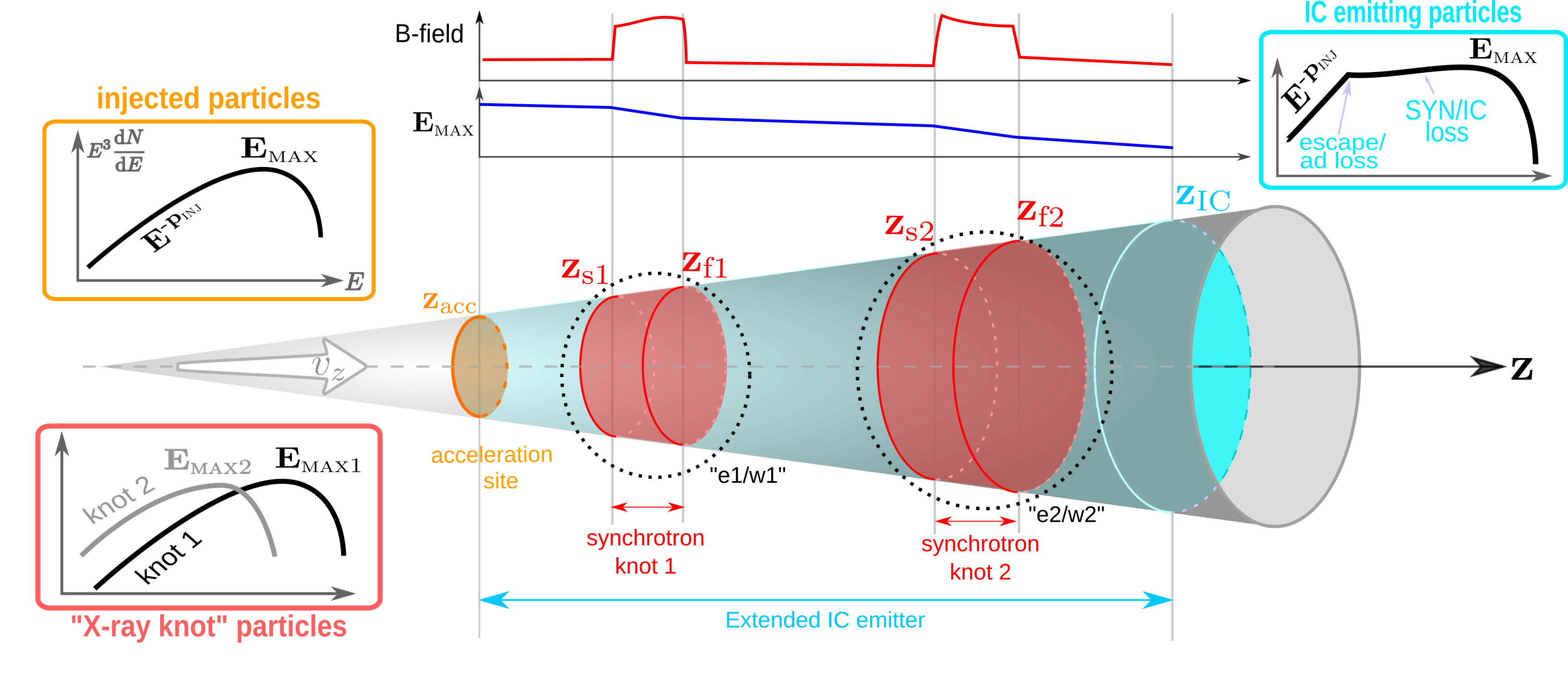}} 
\caption{\label{fig:jet}Schematic illustration of our model for the jet, emission regions, and expected energy distributions of particles.}
\end{figure*}

Recently, the HAWC collaboration has reported the detection of $\gtrsim25$~TeV gamma rays from the jets of SS433 \citep{HAWC18}. The locations of the gamma-ray emission are $\sim30$~pc away from the binary both in the eastern and western side and coincide with nonthermal X-ray emitting regions \citep{Watson83,Yamauchi94,Brinkmann96,SafiHarb97,SafiHarb99}. This indicates that these regions are plausible sites for the acceleration of high-energy particles. SS433/W50 has also been detected with \fer \citep{Bordas15,Bordas17,Xing19,Rasul19,Sun19}, though the origin of the \ac{he} gamma-ray emission ($>100$~MeV) remains unclear. Imaging atmospheric Cherenkov telescopes have not yet detected this system either from the jets nor from the binary in the \ac{vhe} regime ($>100$~GeV) \citep{Veritas17,MAGIC18}. 

Multi-wavelength emission from the jets of SS433 provides us with valuable opportunities to study the acceleration of particles in astrophysical jets in great detail. There are a number of theoretical studies on the nonthermal emission in microquasars \citep[e.g.,][]{Atoyan99,Heinz02,Romero03,Bosch-Ramon06,Gupta06,Orellana07,Reynoso08,Perucho08,Romero08,Bordas09,Vila10,Zdziarski14,Pepe15,Molina18,2018MNRAS.481.1455K,Zhang18,Reynoso19}. The detection of TeV gamma rays from SS433 provides new important constraints on emission models. However, there are only a few studies that utilize new observational data, and the results remain somewhat controversial. \cite{HAWC18} focused on the eastern region and concluded that the radio, X-ray and \ac{vhe} data can be well-fit with leptonic models. On the other hand, \cite{Xing19} studied the western region and argued that leptonic models have difficulties in explaining the radio and X-ray data simultaneously. Because both papers adopt simple models, where particles are continuously injected throughout the source lifetime \citep[$\sim$20~kyr,][]{Zealey80,Goodall11a} and cooled only via radiative loss, a new theoretical study with more detailed physical consideration is needed to uncover the origin of the emissions from the jets of SS433.

Here, we study the nonthermal emission from the SS433 jets in the light of recent multi-wavelength observations. We aim to assess the validity of leptonic models, to examine the efficiency of particle acceleration and processes responsible for that, and to study prospects for future observations. Going beyond prior work noted above, we consider the spatial distribution of emission along the jet and include adiabatic loss due to the jet expansion. 

In Fig.~\ref{fig:jet}, we schematically show how an astrophysical jet and the emission sites can be modeled. Galactic and extragalactic jets often contain multiple compact emitting regions (``knots"), which may appear distinct due to various reasons. For example, an X-ray knot may correspond to the region with a locally enhanced magnetic field. In the case of SS433, the jets are launched to both eastern and western sides, each of which contains multiple X-ray knots, and in Fig.~\ref{fig:jet} we only show one side of the jet. Further description of Fig.~\ref{fig:jet} is presented in Sec.~\ref{subsec:particle_qual}. We mainly analyze the emission from the innermost knots (``e1'' and ``w1'') to compare results with \cite{HAWC18} and \cite{Xing19}, but also address the emission of different regions qualitatively. Also, we consider the case where the acceleration site matches the onset of the innermost knot. We only study leptonic emission, since hadronic emission is already disfavored as the dominant source of TeV gamma rays from SS433~\citep{HAWC18}. However, the inferred electron acceleration efficiency can also have implications for the production of high-energy protons.

In Sec.~\ref{sec:model}, we present a general model for the particle evolution and emission in relativistic jets. In Sec.~\ref{sec:ss433}, we briefly review the observational properties of SS433. In Sec.~\ref{sec:result}, we compare our model predictions with the multi-wavelength data from the two X-ray knots. In Sec.~\ref{sec:size}, we study the morphology of the emission regions. In Sec.~\ref{sec:limitation}, we note limitations to our results. In Sec.~\ref{sec:conclusion}, we summarize our results and discuss further implications.

\section{Physical conditions in relativistic jet} \label{sec:model}

\subsection{Energetics}

Let us consider a relativistic jet of total power $L\jet$. The jet radius depends on the coordinate on the jet axis, denoted as $z$: $R=R(z)$. The jet energy flux is
\be\label{eq:L_k}
L\jet = \pi R^2(z) w\Gamma^2 v_z,
\ee
where $w$ is the plasma enthalpy per unit volume, $v_z$ is the jet velocity, and $\Gamma=\left[1-(v_z/c)^2\right]^{-\nicefrac12}$ is the bulk Lorentz factor. The enthalpy is carried by protons, leptons and magnetic fields. We assume that the jet power is distributed by dimensionless fractions $\vk_p$, $\vk_e$, and $\vk\Bf$ for each component, such that $\vk_p + \vk_e + \vk\Bf=1$. The proton contribution determines the mass flux:
\be
\dot{M}\jet = \pi R^2(z) n_pm_p\Gamma v_z.
\ee

The magnetic field is necessary for the acceleration of particles and the production of synchrotron radiation. Its energy is carried as the Poynting flux:
\be\label{eq:L_b}
L\Bf= \frac{v_z \Gamma^2R^2(z) B^2(z)}{4}, 
\ee
where $B$ is the strength of the magnetic field in the plasma comoving frame, and we assume that it is perpendicular to the jet velocity in the above expression for simplicity. From this, we can express $B$ as
\be\label{eq:b_jet}
B(z)=\frac{2}{\Gamma R(z)}\sqrt{\frac{\vk\Bf L\jet}{v_z}}.
\ee

\subsection{Particle Acceleration}

The process responsible for the acceleration of nonthermal particles in the microquasar jet is not certain. We characterize it by an efficiency $\eta\acc$($>$1). The time required for a particle to gain energy $E$ is
\be \label{eq:t_acc}
\tau\acc=\eta\acc\frac{r\Ge}{c},
\ee
where $r\Ge={E}/{(eB)}$ is the relativistic gyroradius and $e$ is the elementary charge.

The confinement of particles in the acceleration region implies the following condition:
\be
R(z) > \sqrt{6\tau\acc D},
\ee
where $D$ is the particle diffusion coefficient and we assume a three dimensional case. This is similar to the Hillas criterion \citep[$r\Ge$$<$$R$,][]{1984ARA&A..22..425H}. We introduce a parameter $\eta_g$($>$1), known as gyro-factor, and characterize the spatial diffusion as $D=\eta_g D\bohm$, where $D\bohm=cr\Ge/3$ is the Bohm diffusion coefficient. Combining above equations, we obtain
\be
R(z)B(z) > \frac{E}{e}\sqrt{2\eta\acc\eta_g},
\ee
which constrains the maximum energy of particles that can be confined in the jet during the acceleration process:
\be\label{eq:Emax_con}
E<E_{\max}^{\rm con} = \frac e{\Gamma}\sqrt{\frac{2\vk\Bf L\jet}{v_z\eta\acc\eta_g}}.
\ee
The confinement condition is not the only constraint, as the particle acceleration is also limited by energy losses. In this work we consider emission from electrons. Thus, in the highest-energy regime, the synchrotron cooling may provide the dominant loss mechanism. The synchrotron cooling time is
\be
\tau\syn = \frac{3(m_ec^2)^2}{4c\sigma_TE}\left(\frac{B^2}{8\pi}\right)^{-1},
\ee
where $\sigma_T$ is the Thomson cross section and $m_e$ is the electron mass. The acceleration is possible while it proceeds on a shorter timescale than cooling, $\tau\acc<\tau\syn$, which sets the maximum energy of particles:
\be\label{eq:Emax_syn}
    E<E_{\max}^{\rm syn} = m_ec^2\sqrt{\frac{6\pi e}{\sigma_T\eta\acc B}}\,.
\ee
The magnetic field used above should be evaluated in the acceleration site, which can in principle differ from that in the emission region (see Fig~\ref{fig:jet}). 

\subsection{Particle Cooling}

Accelerated particles are subject to energy losses due to adiabatic and radiative cooling. The adiabatic loss rate due to the expansion of the jet is 

\be \label{eq:ad_cool}
\dot{\gamma}\ad = \frac{\gamma}{3}\frac{d\ln \rho}{d\tau} = -\frac{2}{3}\frac{\Gamma v_z}{R(z)}\frac{\partial R}{\partial z}\gamma,
\ee
where $\rho$ is number density of matter in the jet, and we assume that the jet speed $v_z$ is constant (thus $\rho R^2$ is also constant). 

The radiative losses for high-energy electrons are dominated by the synchrotron emission and \ac{ic} scattering. The synchrotron loss rate is

\be
  \dot{\gamma}\syn = -\frac{4c\sigma_T\gamma^2}{3m_ec^2}\left(\frac{B_{\rm eff}^2}{8\pi}\right),
\ee
where the magnetic field $B_{\rm eff}$ corresponds to the averaged effective field strength. If the magnetic field strength is constant, $B$, and the pitch angles between the particle velocity and the magnetic field are random, we should use $B_{\rm eff}^2=2B^2/3$. In more general cases, the field strength may have spatial variation within the emission region probably due to magnetic turbulence \citep[e.g.,][]{Bykov12,Kelner13,2019arXiv190711663D}. Then, the magnetic field distribution function is needed to obtain $B_{\rm eff}$.

To describe the \ac{ic} losses we need to consider the contributions from all relevant photon fields. A precise treatment requires integration over photon energy and angular distribution, which can be complex. Fortunately, the photon energy distribution is often described by a black-body like spectrum, where the photon field is defined by its temperature $T$ and energy density $u_{\rm rad}$ or, equivalently, by the dilution coefficient: 
\be
\kappa = \frac{15\hbar^3c^3u_{\rm rad}}{\pi^2k_B^4T^4},
\ee
where $k_B$ is the Boltzmann constant and $\hbar = h/(2\pi)$ is the Dirac constant. If the jet bulk Lorentz factor is small and the target photon is black-body like, the simple approximate description obtained by \citet{Khangulyan14} is applicable for the energy losses including Klein-Nishina effect:
\be\label{eq:kn_losses}
\dot{\gamma}\ic =- \frac{3\sigma_Tk_B^2T^2m_ec^2\kappa}{4\pi^2\hbar^3c}G_{\rm iso}^{(0)}\left(4\gamma \frac{k_BT}{m_ec^2}\right)g_{\rm iso}\left(4\gamma \frac{k_BT}{m_ec^2}\right)\,,
\ee
where the function $G_{\rm iso}^{(0)}(u)$ and $g_{\rm iso}(u)$ are given in their Eq.~(38) and (20) respectively. In more general cases, when the bulk Lorentz factor is large or the photon direction deviates from isotropic, we need to perform integration over the photon angular distribution.

\subsection{Particle Evolution}

The distribution of nonthermal particles in the jet can be described with the energy-spatial density, $n$, as $dN = n d\gamma dz$, where $dN$ is the number of particles. The density is described by the relativistic transport equation \citep[see][for detail]{1989ApJ...340.1112W,2018ApJ...865..144V}:
\be
  \begin{split}
    \Gamma\left( \frac{\partial}{\partial t} + v_z\frac{\partial}{\partial z}\right)n(\gamma,t,z)
   +\frac{\partial}{\partial\gamma}[\dot{\gamma}_{\rm cool}(&\gamma,t,z)n(\gamma,t,z)]\\ 
   &= \dot{q}\inj(\gamma,t,z),
  \end{split}
\ee
where $\dot{q}\inj dz$ is the rate of particle injection in the jet segment $(z,z+dz)$. We assume that nonthermal particles are accelerated at a specific coordinate $z=z\acc$:
\be
\dot{q}\inj(\gamma,t,z)=\dot{q}\sub{0}(\gamma,t)\delta_D(z-z\acc)\,
\ee
where $\delta_D$ is the Dirac function. {For simplicity, we consider the case where the cooling rate $\dot{\gamma}_{\rm cool}$ depends only on $\gamma$. Then, the solution of the transport equation is obtained analytically:}
\be\label{eq:density_local}
n(\gamma,z,t) = \frac{\dot{\gamma}_{\rm cool}(\tilde{\gamma})}{\Gamma v_z\dot{\gamma}_{\rm cool}(\gamma)}\dot{q}\sub{0}(\tilde{\gamma},\tilde{t})\mathcal{H}(z-z\acc),
\ee
where $\tilde{t}$~=~$t-(z-z\acc)/v_z$, $\mathcal{H}$ is the Heaviside function and $\tilde\gamma$ is an energy parameter determined by 
\be \label{eq:gamma_i}
z - z\acc = \Gamma v_z \int_{\gamma}^{\tilde{\gamma}} \frac{d\hat{\gamma}}{-\dot{\gamma}_{\rm cool}(\hat{\gamma})}.
\ee
To calculate the total emission from a specific region along the jet, we integrate the particle distribution over the emitting region ($z_s$$<$$z$$<$$z_f$):
\be \label{eq:dNdgamma}
\frac{dN}{d\gamma}=\int_{z_s}^{z_f}dz\, n(\gamma,z,t).
\ee
Assuming a steady injection ($\partial \dot{q}\sub{0}/\partial t=0$), we obtain
\be \label{eq:electron}
    \frac{dN}{d\gamma} = \frac{1}{\Gamma v_z\dot{\gamma}_{\rm cool}(\gamma)}\int_{z_s}^{z_f} dz
    \dot{\gamma}_{\rm cool}(\tilde{\gamma})
    \dot{q}\sub{0}(\tilde{\gamma}),
\ee
where $\tilde{\gamma}$ is determined by $z\acc$, $z$, and $\gamma$ via Eq.~(\ref{eq:gamma_i}). Since we focus on compact knots much smaller than the jet length, $(z_f-z_s)\ll z_f$, we assume that the change in radius is also small, $R(z_f) - R(z_s) \ll R(z_f)$, and evaluate $\dot{\gamma}\ad$ and $B(z)$ at $z_I=(z_s + z_f)/2$ to omit the $z$ dependence. Also, we assume that the onset of emission region matches the acceleration site, i.e., $z\acc = z_s$.

We assume that particles are accelerated to a power-law energy distribution above $\gamma>\gamma_{\min}$ with an exponential cutoff:
\be
\dot{q}\sub{0}(\gamma)\propto\gamma^{-p\inj}\exp{(-\gamma/\gamma_{\max})}\mathcal{H}(\gamma-\gamma_{\min}),
\ee
where $\gamma_{\max}$ is defined from either Eq.(\ref{eq:Emax_con}) or (\ref{eq:Emax_syn}).

The power carried by relativistic electrons, $\vk_e L\jet$, defines the normalization for the energy distribution:

\be
\int_{\gamma_{\min}}^{\gamma_{\max}} \dot{q}\sub{0}(\gamma)\gamma d\gamma = \frac{\vk_e L\jet}{m_ec^2\Gamma^2}.
\ee
The value of $\gamma_{\min}$ is theoretically related to the energy scale where thermal particles are injected into the acceleration processes. This is extensively studied but still contains large uncertainties~\citep{Amano12}. We assume a minimum electron energy of 1~GeV. Smaller values of $\gamma_{\min}$ would increase the total electron energy required in the spectral fitting but do not alter the shape in the energy range of interest.

Once the electron distribution is determined, we calculate the spectral energy distribution from synchrotron and \ac{ic} radiation in jet frame, taking into account Klein-Nishina effect \citep[e.g.,][]{BG70,Aharonian10,Khangulyan14}. In more general cases when the bulk Lorentz factor is large, we need to apply relativistic transformations to obtain the spectral energy distribution in the observer frame.

\subsection{Qualitative Description of the Particle Spectrum}
\label{subsec:particle_qual}

If the particle cooling time is shorter than the advection time, \(\tau_{\rm adv}=(z_{f}-z_{s})/(\Gamma v_{z})\), the particle spectrum is described by the standard formula (fast cooling regime):
\be
\frac{dN}{d\gamma} \simeq \frac1{\dot{\gamma}(\gamma)}\int_{\gamma}^{\infty}d\tilde{\gamma}\dot{q}(\tilde{\gamma}).
\ee
For a power-law injection with $p\inj\simeq2$, this expression reduces to $dN/d\gamma\simeq\dot{q}(\gamma)\tau_{\rm cool}$. In this regime, the electron energy distribution has a break, at which the particle power-law index is changed by 1, caused by the transition from the synchrotron/Thomson to adiabatic cooling (or un-cooled). This is qualitatively shown in Fig.~\ref{fig:jet} labeled as ``IC emitting particles." 

If we consider emission from compact knots, the advection time may be shorter than the cooling time (slow cooling regime). The particle spectrum is described by
\be
\frac{dN}{d\gamma} \simeq\dot{q}(\gamma)\tau_{\rm adv},
\ee
which has a shape similar to the injection spectrum. This is qualitatively shown in Fig.~\ref{fig:jet} labeled as ``knot 1." In knots further away from the acceleration site, the particle number per unit energy per unit volume remains unchanged at lower energies. However, the cutoff energy in the spectrum may be reduced due to the cooling; this can be directly seen from Eq.~(\ref{eq:gamma_i}). This is qualitatively shown in Fig.~\ref{fig:jet} labeled as ``knot 2." 

\subsection{Knot Size}

Observations in radio, optical, or X-ray often reveal knots in Galactic and extragalactic jets. In general, their size may be determined by either of the following factors: (i) nonthermal cooling; (ii) size of the jet region with an enhanced magnetic field; (iii) size of the region where the acceleration takes place or time elapsed since the onset of the acceleration process. 

If the knot size is determined by the particle cooling, the energy requirements for the acceleration process are minimal and the spectral slopes are typical ones for the fast cooling regime. If the particle acceleration occurs at a specific location in the jet, advective particle transport determines the knot size, $s$, as $s \simeq \Gamma v_z \tau_{\rm cool}(E)$. If synchrotron losses are dominant, the cooling time depends on the particle energy, $\tau_{\rm cool}\propto 1/E$. The synchrotron emission frequency $\omega$ and particle energy relate as $E\propto \sqrt{\omega / B}$, and thus the knot size should depend on the photon frequency as $s \propto 1/\sqrt{\omega}$. If adiabatic losses dominate the particle cooling, the knot size does not depend on the particle energy. Adiabatic cooling generally does not produce compact knots, except for specific hydrodynamic structures of the jet. For example, for a constant velocity jet and conical or parabolic shape, it operates on a scale comparable to the jet length. 

The synchrotron emissivity is sensitive to the magnetic field strength. If some portion of the jet has an enhanced magnetic field, it may appear as a compact, bright spot. This may result in different morphology for the synchrotron and \ac{ic} emission (see Fig.~\ref{fig:jet}).

The acceleration does not necessarily proceed at a specific point in the laboratory frame, and may be associated with a fluid element. In this case, the knot size depends on the size of the acceleration site and the typical diffusion length, $\lambda_{\rm D}$. Because $\lambda_{\rm D}\propto \sqrt{\tau}$, the size of the knot should have a weaker dependence on the photon frequency as compared to the synchrotron cooling scenario. 

There can also be a possibility that the acceleration has started recently and the knot size is limited by the advection distance since the moment of onset of the acceleration. This would produce a gradual increase in the knot size with time. However, this may be difficult to observe on a reasonable timescale. 

\section{Application to SS433} \label{sec:ss433}

\subsection{Properties of SS433 Jets}
\label{sec:jet}

Observations of the jets of SS433 provide necessary parameters for the formalism presented in Sec.~\ref{sec:model}. We adopt a distance of $d$~=~5.5~kpc, which is obtained from deep radio imaging \citep{Blundell04}. Long-term observations and kinematic modelings of the Doppler-shifted emission lines place tight constraints on the jet precession model. They yield a jet speed of $v_z=0.26c$ \citep{Margon89,Eikenberry01}, or equivalently, $\Gamma = 1.04$. Because this is only mildly-relativistic, we do not take relativistic effects into account. 
Models of the jet emission indicate that the mass-loss rate at the jet base is $\dot{M}\jet\gtrsim 10^{-7}M_\odot$~yr$^{-1}$, which leads to the estimates for the kinetic energy without rest mass energy, $(\Gamma-1)\dot{M}\jet c^2$, which typically fall within $\sim(0.2-5)\times10^{39}$~erg~s$^{-1}$. \citep[e.g.,][]{Kotani96,Brinkmann00,Marshall02,Brinkmann05,Medvedev10,Waisberg19}. Because estimates for the total jet power $L\jet$ have uncertainties, instead of using $\vk_e$ and $\vk\Bf$, we will leave $L_e(=\vk_e L\jet)$ and $L_B(=\vk\Bf L\jet)$ as free parameters. We adopt the jet kinetic energy of $10^{39}$~erg~s$^{-1}$ at the jet base, and assume that part of this is dissipated to $L_e$ and $L_B$, i.e., we keep $L_e + L_B < 10^{39}\ {\rm erg\ s^{-1}}$.

We assume a conical jet, and parametrize the radius with the opening angle $\alpha_j$ as $R(z)=z\alpha_j$. We adopt a radius of $R(z_I)=6$~pc, comparable to the size of X-ray emission~\citep{SafiHarb97}. With this parameterization, Eq. (\ref{eq:ad_cool}) reduces to 
\be \label{eq:ad_cool_conical}
-\dot{\gamma}\ad = \frac{2}{3}\frac{\Gamma v_z}{z}\gamma.
\ee

The photon field is also a necessary ingredient as a target for the \ac{ic} scattering. We adopt a Galactic radiation field composed of the cosmic microwave background ($T,u_{\rm rad}$)=(2.7~K, 0.26~eV~cm$^{-3}$), far-infrared (30~K, 0.6~eV~cm$^{-3}$), optical/near-infrared (5000~K, 0.6~eV~cm$^{-3}$) and ultraviolet (20000~K, 0.6~eV~cm$^{-3}$) photons \citep{Porter17,Popescu17}. 

The accretion disk in SS433 has a high bolometric luminosity of $L_{\rm bol}\simeq 10^{40}$~erg~s$^{-1}$ and temperature of $T\simeq 10^5$~K~\citep[e.g.,][]{Antokhina87,Begelman06}. At the knot regions, the energy density of this disk emission is $u_{\rm rad}\sim2~{\rm eV~cm}^{-3}$. However, due to Klein-Nishina effect, the contribution of this component to the \ac{ic} spectrum is suppressed above $E_\gamma \gtrsim$~10~GeV. Furthermore, even in the Thomson regime, the emissivity of \ac{ic} emission scales as $L_{\rm IC}\propto u_{\rm rad}T^{(\alpha_e-3)/2}$, where $\alpha_e$ is the spectral index for the electron distribution~\citep{BG70,Aharonian97}. As we focus on the knot emission, where electrons are in slow-cooling regime and have a hard spectrum, the contribution of the disk photons is subdominant compared to the cosmic microwave background. Thus in what follows we do not consider this component. We have verified that this emission contributes negligibly to the detected GeV emission unless we adopt unrealistically high energy density of $u_{\rm rad}\sim 100~{\rm eV~cm}^{-3}$.

\subsection{Multi-wavelength Observations Toward X-ray Knots}
\label{sec:data}

The jets from SS433 have been intensively studied with multi-wavelength observations. Based on {\it ROSAT} and {\it ASCA} X-ray data, \citet{SafiHarb97} defined distinct circular regions to east (e1, e2, e3) and west (w1, w2) from the binary. Combining {\it RXTE} data, the emission from e1 is fit with a single power law of $\Gamma_{\rm ph}=1.43\pm 0.1$~\citep{SafiHarb99}, while e2 is a broken power-law spectrum of $\Gamma_{\rm ph,1}=1.6^{+0.2}_{-0.3}$ and $\Gamma_{\rm ph,2}=2.6^{+0.6}_{-0.3}$ with a break at $E_b=3.0^{+0.6}_{-0.5}$~keV~\citep{SafiHarb97}. The eastern jet is also observed with {\it XMM-Newton} by \citet{Brinkmann07}. They found $\Gamma_{\rm ph}=2.17\pm0.02$ for the brightest region in the eastern jet and $\Gamma_{\rm ph}=1.85\pm0.06$ for a region closer to the binary. These regions are not identical to e2 and e1, though they overlap. It should be noted that the derived Galactic column density in \citet{SafiHarb97} is $N_{\rm H}=1.2^{+0.8}_{-0.5}\times10^{21}~{\rm cm^{-2}}$ for e2, while it is $N_{\rm H}=5.6^{+0.1}_{-0.1}\times10^{21}~{\rm cm^{-2}}$ in \citet{Brinkmann07}. This may cause differences in the derived photon index.

In the \ac{vhe} regime, the H.E.S.S., MAGIC, and VERITAS collaborations placed upper limits on the flux from knots (e1, e2, w1, w2) and termination region (e3)~\citep{Veritas17,MAGIC18} following the definitions in \citet{SafiHarb97}. The HAWC collaboration reported the detection of \ac{vhe} photons ($\gtrsim$ 25~TeV) from regions that coincide with X-ray knots. The eastern emission is seemingly radiated from a region spanning over e1 and e2, and the western component is likely centered at w1, though both are not yet well localized.

In the radio band, fluxes from the knot regions are uncertain. The termination region (e3) is prominent in radio images and well-correlated with X-ray intensity maps. However, the knots, e1, e2, and w1, are not resolved in the 2.7~GHz map by the Effelsberg telescope \citep{Geldzahler80} nor in the recent 150~MHz map by LOFAR \citep{LOFAR18}. This suggests that the contribution from X-ray knots to the observed radio intensity may be sub-dominant, and the radio fluxes should be treated as upper limits. Radio spectral index measurements would provide useful constraints on the spectral shape of nonthermal electrons. \cite{Downes86} produced a radio spectral index map utilizing 1.7, 2.7 and 4.75~GHz data. However, the X-ray knots are not well localized also in this map.

In contrast to other wavelengths, recent results in the \ac{he} regime are controversial. \citet{Bordas15} suggests that emission from nonthermal protons accelerated in the jet termination shock best explains the emission detected with \fer. The analysis by \citet{Xing19} suggests a one-sided jet morphology toward the w1 region. While these papers indicate no signature of time variation, \citet{Rasul19} reports $\sim 3\sigma$ evidence for temporal modulation of the gamma-ray emission with the precession period of the jet, which would indicate core origin \citep[see also][]{Molina18}. \citet{Sun19} suggests that the morphology of the GeV emission is consistent with originating from the radio nebula W50. The spectrum and morphology are somewhat different from each other. Thus, it is difficult at this point to clearly define the \ac{he} gamma-ray flux from the X-ray knots. 

Further observations are needed to quantify the multi-wavelength properties of the X-ray knots better. Here, we constrain our model parameters by using the same dataset for radio, X-ray and \ac{vhe} emission as in \cite{HAWC18} and \cite{Xing19}, aiming at comparing model predictions with them. We also compare our model spectra with the GeV data from \cite{Bordas17,Xing19,Rasul19,Sun19}, which are not used in the model fitting.

We adopt the definition of e1 as a circular region of radius $3.5'$ centered at $24'$ east from SS433, and w1 a circle of radius $3.75'$ centered at $19'$ west. These translate into parameter ($z_s$,$z_f$) as (32~pc, 44~pc) for e1 and (24~pc, 36~pc) for w1 in Eq.~(\ref{eq:dNdgamma}). We note that the {\it XMM-Newton} data used in \citet{HAWC18} are taken from a slightly larger region (a circle of $6'$ radius centered at e1), which we do not take into account here. As we consider the emission from a region that spans approximately 10~pc across, we do not expect any influence from the orbital or precession phase, which may appear only on a significantly smaller scales.

\section{Nonthermal leptonic emission from knots in SS433 jets} \label{sec:result}

\begin{center}
\begin{figure}[t]
\includegraphics[width=\columnwidth]{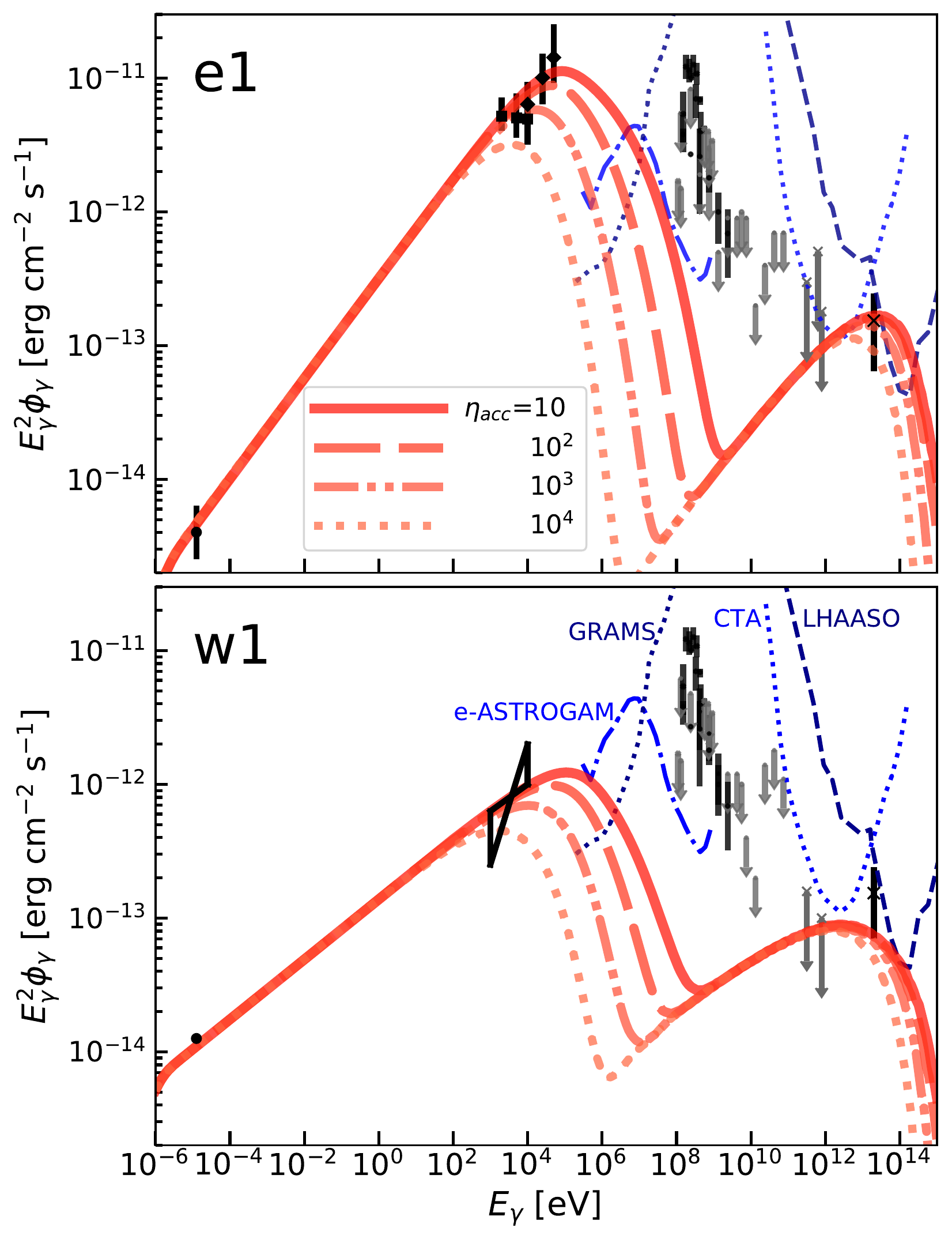} 
\caption{Broad-band spectral energy distribution of the e1 (top) and w1 (bottom) region. Orange curves are model predictions for different choices of $\eta\acc$, as labelled. Black and gray points are observational data and upper limits, respectively, from \cite{Geldzahler80} (radio), \cite{Brinkmann07,SafiHarb97,SafiHarb99} (X-ray), \cite{Bordas17,Xing19,Rasul19,Sun19} (\ac{he}), \cite{MAGIC18,Veritas17,HAWC18} (\ac{vhe}). Expected sensitivities are also shown for CTA \citep[North, 50 h;][]{CTAbook}, LHAASO \citep[1 yr;][]{LHAASObook}, {\it e-ASTROGAM} \citep[3 yr;][]{DeAngelis2017} and {\it GRAMS} \citep[3 yr;][]{Aramaki2020}.}
    \label{fig:sed}
\end{figure}
\end{center}

In Fig.~\ref{fig:sed}, we show the spectral energy distribution for the e1 and w1 region. Our leptonic models explain the radio, X-ray and \ac{vhe} data. For the GeV data, our predictions in the HE regime are far below the data for both regions. This indicates that it is difficult to explain the GeV data simultaneously with other wavelength data in the framework of our leptonic models from knot regions. Thus, most GeV photons should be produced in different regions or by different mechanisms.

In Table~\ref{table:bestfit}, we list the required parameters for the fit. The slope $p\inj$ is determined from the radio and X-ray data, while $L_B$ and $L_e$ are derived by combining them with the HAWC data. The derived magnetic field strengths are $16\ {\rm \upmu G}$ and $9\ {\rm \upmu G}$ for e1 and w1, respectively. 

The mechanism responsible for the maximum energy cannot be determined from this fit. We temporarily focus on the case where it is limited by synchrotron losses (Eq.~\ref{eq:Emax_syn}). Then, the magnetic field and acceleration efficiency, $\eta\acc$, define the maximum electron energy: 
\be
E_{e,\max}^{\rm syn} = 1.5\ {\rm PeV}\left(\frac{\eta\acc}{10^2}\right)^{-\nicefrac{1}{2}}\left(\frac{B}{\rm 16~\upmu G}\right)^{-\nicefrac{1}{2}}.
\ee

\begin{table}[b]
\begin{center}
     \caption{Model parameters}
\begin{tabular}{cccc} \hline
     Region & $p\inj$ & $L_e$ [10$^{39}$ erg~s$^{-1}$] & $L_B$ [10$^{39}$ erg~s$^{-1}$] \\ \hline
     e1 & 2.25 & 0.02 & 0.18 \\ 
     w1 & 2.55 & 0.08 & 0.06 \\ \hline
\end{tabular}
     \label{table:bestfit}
\end{center}
\end{table}

In our model, the hard X-ray data require $\eta\acc \lesssim 10^2$ for both regions. Although our model does not specify the acceleration processes, it would be helpful to interpret $\eta\acc$ in the framework of two representative scenarios. First, we consider diffusive shock acceleration. In this mechanism, particles gain energy as they cycle upstream and downstream across the shock front. The acceleration timescale in a parallel shock is given by $\tau\acc^{\rm DSA} \simeq 10D/v_{\rm sh}^2$ \citep[e.g.,][]{Bell13}. This translates into the efficiency in Eq.~(\ref{eq:t_acc}) as 
\be \label{eq:DSA}
\eta\acc^{\rm DSA} \simeq \frac{10\eta_g}{3(v_{\rm sh}/c)^2} \simeq 10^2\left(\frac{\eta_g}{2}\right)\left(\frac{v_{\rm sh}}{0.26c}\right)^{-2}.
\ee
Thus, our results suggest that the diffusion coefficient may satisfy $\eta_g\lesssim 2$, indicating strong particle confinement close to the Bohm limit. Such a high particle acceleration efficiency is known to be achieved in young supernova remnants \citep[e.g.,][]{Stage06,Uchiyama2007,Tsuji19}, while it is thought to be much more inefficient in extragalactic black hole jets \citep[e.g.,][]{Araudo15,Inoue2016,Tanada2019} possibly due to the inefficiency of the diffusive shock acceleration mechanism in the relativistic regime \citep{Bell18}. 

Second, we consider the stochastic acceleration. In this mechanism, particles gain energy as they are resonantly scattered by magnetohydrodynamic turbulence \citep[e.g.,][]{Dermer09}. Assuming that the smallest turbulence wavenumber is equal to $R^{-1}$, the timescale for acceleration is given by
\be \label{eq:Stochastic}
\tau\acc^{\rm S} \simeq \frac{1}{\kappa_B}\left(\frac{v_A}{c}\right)^{-2}\left(\frac{r_L}{R}\right)^{2-q}\tau_{\rm dyn},
\ee
where $\kappa_B$ is the ratio of the strength of turbulent field compared to the background field, $\tau_{\rm dyn}=R/c$ is the dynamical timescale and $q$ describes the spectrum of the turbulence. This expression is derived under quasi-linear approximation ($\kappa_B\ll1$), but has a wider applicability \citep{OSullivan09}. The Alfven velocity $v_A=B/\sqrt{4\pi m_pn_p}$ can be expressed in the form of 
\be
\frac{v_A}{c} = \sqrt{\frac{L_B}{\Gamma\dot{M}c^2}}
\ee
Combining the above equations and assuming the Kolmogorov-type spectrum ($q=5/3$), we have
\be
\eta\acc^{\rm S}\simeq \frac{10^3}{\kappa_B}\left(\frac{\dot{M}}{10^{-7}M_\odot~{\rm yr^{-1}}}\right)\left(\frac{L_B}{10^{38}\ {\rm erg~s^{-1}}}\right)^{-1}.
\ee
for $R=6$~pc and $E=1$~PeV. Thus, the stochastic acceleration is likely insufficient to reach the high efficiency of $\eta\acc < 10^2$, though it is not firmly ruled out due to simplifications in this estimate.

So far we have focused on the case where electron energy is limited by the synchrotron loss. If escape is efficient, the confinement limit should be dominant for electrons when $E_{e,\max}^{\rm con} < E_{e,\max}^{\rm syn}$, or,
\be \label{eq:confinement_case}
\eta_g > 20 \left(\frac{R}{6\ {\rm pc}}\right)^{2}\left(\frac{B}{16\ \upmu{\rm G}}\right)^{3}.
\ee
In this case, our results constrain the product $\eta\acc \eta_g$ to $\eta\acc \eta_g\lesssim 10^3$. Combining this with Eq.~(\ref{eq:confinement_case}), we obtain $\eta\acc\lesssim 10^2$.

The acceleration of leptons may imply the presence of nonthermal protons because they have a larger Larmor radius and are more easily injected into the acceleration processes. Synchrotron losses are inefficient for protons, and the maximum energy is limited by confinement:
\be
E_{p,\max}^{\rm con} = 5\ {\rm PeV}\left(\frac{\eta\acc\eta_{g}}{10^2}\right)^{-\nicefrac{1}{2}}\left(\frac{L_B}{10^{38}\ {\rm erg~s^{-1}}}\right)^{\nicefrac{1}{2}}.
\ee
If we assume that the same acceleration process is at work both for electrons and protons, we can apply the same value of $\eta\acc$. Then, the constraint $\eta\acc \lesssim 10^2$ formally suggests that SS433 can accelerate protons beyond a PeV, if the Bohm factor $\eta_g$ is sufficiently small. 

As noted in Sec.~\ref{sec:data}, the radio flux may be dominated by other components. If we treat the radio data as upper limits, the parameter $p\inj$ can become as small as 1.9. The spectral turnover and cutoff predicted in our model and constraints on $\eta\acc$ remain unchanged because they are derived from the hard X-ray data and determined by the timescales of synchrotron cooling $\tau\syn$ and transport $\tau_{\rm adv}$.

\subsection{Comparison with Previous Studies}

\begin{center}
\begin{figure}[t]
    \includegraphics[width=\columnwidth]{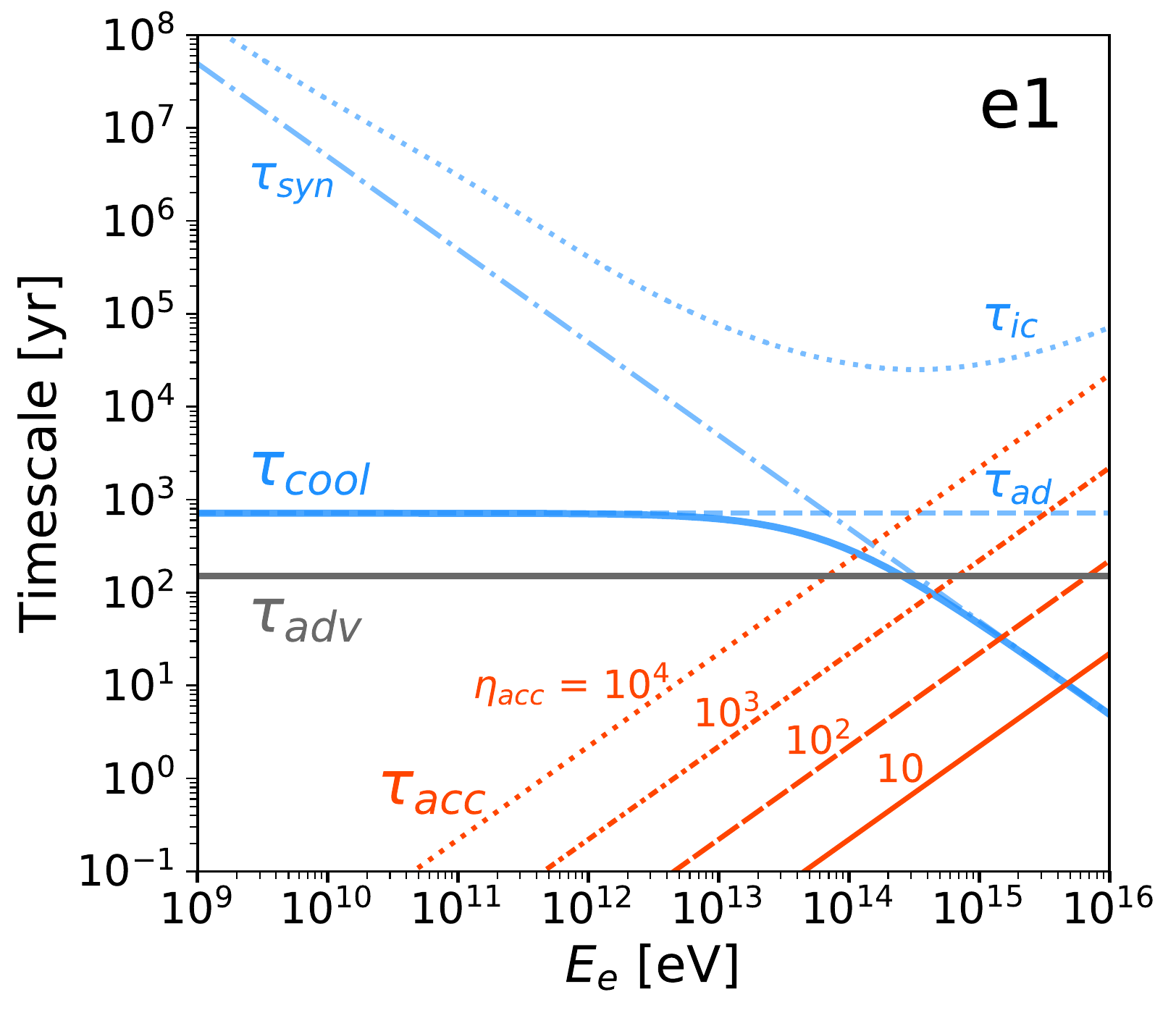}
    \caption{Cooling timescales ($\tau = \gamma/|\dot{\gamma}|$) for different processes for the e1 region, as marked. For comparison, timescales of confinement and acceleration are also shown.}
   \label{fig:cool}
\end{figure}
\end{center}

We now compare our results with other recent studies. Our model spectra are significantly different from results by \cite{HAWC18} (e1) and \cite{Xing19} (w1). They require a hard spectrum of $p\inj$~=~$1.9$ for electrons, while we derive $p\inj$~=~2.25 and 2.55 for e1 and w1, respectively. The main differences are twofold. First, they calculate the evolution of particles assuming continuous injection throughout the source lifetime, for which they adopted $t_{\rm lifetime}\simeq$30~kyr. Thus, their spectra show cooling breaks in the electron spectra at $E_e$~=~$2~(B/16\ {\rm \upmu G})^{-2}(t_{\rm lifetime}/30\ {\rm kyr})^{-1}$~TeV, and require a hard $p\inj$. In contrast, we integrate the particle spectrum from $z_s$ and $z_f$~(Eq.~\ref{eq:dNdgamma}), and the effective lifetime is set by $\tau_{\rm adv}\simeq(z_f-z_s)/v_z\simeq$150~yr. Second, while they only include radiative losses, we also consider adiabatic loss. In Fig.~\ref{fig:cool}, we compare energy loss timescales for different processes. Adiabatic losses dominate below 100~TeV, significantly limiting the total electron energy. Note that we employ a simple case of a conical jet to evaluate the adiabatic loss. The jets may be collimated by surrounding material and keep nearly cylindrical ($\partial R/\partial z = 0$), experiencing no adiabatic loss. In our calculation, since the effect of particle transport is dominant over that of adiabatic cooling ($\tau_{\rm adv}<\tau\ad$), the results remain unchanged for different modeling of jet expansion, as long as we focus on the emission from the synchrotron knots.\\

\subsection{Prospects for Future Observations}

We examine expectations for future observations. The hard X-ray (10--100~keV) observations will be most critical. {\it NuSTAR} can provide a better determination of the spectrum from both regions in this regime. In the MeV--GeV band, planned telescopes such as {\it GRAMS} \citep{Aramaki2020}, {\it e-ASTROGAM }\citep{DeAngelis2017}, and AMEGO \citep{AMEGO2019} will be able to study the highest energy synchrotron photons, though the localization of emitting region would be difficult for their expected angular resolutions. Our models predict that these observations would detect spectral turnover and cutoff~(Fig.~\ref{fig:sed}), placing strong constraints on physical properties and acceleration processes. In the \ac{vhe} regime, our results indicate that CTA and LHAASO observations might be able to detect gamma rays from both regions \citep{CTAbook,LHAASObook}.

\section{Morphology of Emission Regions} \label{sec:size}

The location of emission sites is an important ingredient in our model. In this section, we examine explanations for the size of X-ray knots and briefly discuss uncertainties due to different morphology of the X-ray and \ac{ic} emission.

\subsection{X-ray Knot Size in SS433 Jets} \label{subsec:knot}

The X-ray images in the $\sim$~1--10 keV range show a clear feature of knots with a size comparable to the e1 region, $s^{\rm X}\simeq5$--10~pc~\citep{SafiHarb97}. The typical energy of electrons responsible for the X-ray emission at 1~keV is $30~({B}/{15\rm\,\upmu G})^{-1/2}$~TeV, and the synchrotron cooling time for these electrons is $1.5~({B}/{15\rm\,\upmu G})^{-3/2}$~kyr. The advection length during that is 
\be
s\syn^{\rm X}=120~\left(\frac{B}{15\rm\,\upmu G}\right)^{-\nicefrac{3}{2}}\left(\frac{v_z}{0.26c}\right)\rm\,pc\,.
\ee
Thus, with our standard value of $v_z=0.26c$, the X-ray knot size cannot be explained by the synchrotron cooling. Below, we will examine several possible physical processes that may determine the X-ray knot size.

{\bf Synchrotron Cooling: Unlikely.} The knot length may be explained by the synchrotron cooling if we adopt a velocity smaller than $0.26c$, but the spectrum places tight constraints; the emitting particles should be in the fast-cooling regime, where photon spectrum would be $\Gamma_{\rm ph}=2$ (for an $E_e^{-2}$ injection). This contradicts with the hard X-ray spectrum of $\Gamma_{\rm ph}$~$\simeq$~1.5. Thus, synchrotron cooling cannot be responsible for determining the X-ray knot size, unless the electrons are injected with an extremely hard spectrum of $\simeq E_e^{-1}$.

{\bf Adiabatic Cooling: Possible for a non-conical jet.} If the jet is conical, the adiabatic cooling operates on a scale of the jet length, and the knot size would be $s\ad\simeq 3z/2 \simeq 60$~pc, much larger than observed. However, it may experience local expansion or compression due to the pressure from the surrounding material, producing standing coherent waves (called the Mach disk) or more complicated hydrodynamical structures. If the jet has a structure that enhances the adiabatic cooling rate locally, the X-ray knot size could be explained. 

{\bf Enhanced Magnetic Field: Possible.} The magnetic field in the jets may not be distributed uniformly, but have local amplifications probably due to the local compression of the jet or plasma instabilities. The size of the X-ray knot may correspond to the region with an enhanced magnetic field, probably due to turbulence. 

{\bf Very recent acceleration: Unlikely.} There is a theoretical possibility that acceleration has started very recently, $\Delta t$ years ago, and advection determines the size of knots. We cannot rule out the possibility that $\Delta t$ is close to $\tau_{\rm adv}$, but this requires a coincidence. There could also be a possibility that the acceleration takes place in an extended region, rather than at a specific location in the jet, and $\Delta t\ll \tau_{\rm adv}$. However, if this is the case, the injection power required to produce the observed X-ray luminosity would be much larger than in other scenarios. The dissipation of such a large amount of power would have to produce much brighter thermal bremsstrahlung emission from the heated plasma, which is not observed. Thus, this scenario is unlikely.

Future X-ray observations with high angular resolution would be important to distinguish these scenarios. If the knot size is defined by the adiabatic cooling, we should see no dependence on the photon energies. If the knot size corresponds to the size of the region where the magnetic field is enhanced due to turbulence, we expect patchy bright emission inside the emitting region due to the inhomogeneity of magnetic field strength. 

\subsection{IC Emitter Size in SS433 Jets} \label{subsec:ic}

The size of gamma-ray emitting regions, $s^{\rm VHE}$, is not yet clear. The gamma rays with an energy of 25~TeV are predominately generated on the cosmic microwave background, and the emitting electrons have an energy of 100~TeV. The synchrotron cooling time is $0.5~({B}/{15\rm\,\upmu G})^{-3/2}$~kyr, during which these electrons are advected to a distance of 
\be
s\syn^{\rm VHE}=40~\left(\frac{B}{15\rm\,\upmu G}\right)^{-\nicefrac{3}{2}}\left(\frac{v_z}{0.26c}\right)\rm\,pc\,.
\ee
Adiabatic losses may produce a comparable advection distance for a conical jet, or smaller distance if they are locally enhanced. In any case, the IC emitter is likely larger than X-ray knots.

The difference between $s^{\rm VHE}$ and $s^{\rm X}$ can induce uncertainties in our calculation. In particular, though we have used the observed \ac{vhe} flux to derive physical parameter for the e1 and w1 regions, the real TeV flux from these two knots are likely smaller, provided that $s^{\rm VHE} > s^{\rm X}$. This should primarily affect the estimate on $B$. Because the dominant target for \ac{ic} scattering is provided by the diffuse background, synchrotron and \ac{ic} luminosities relate as $L\syn/L\ic \propto B^2$, and thus the magnetic field strength may be larger approximately by a factor of $\sqrt{s^{\rm VHE} / s^{\rm X}}$ when we take the size of emitting regions into account. Future CTA observations would better constrain the size of the \ac{ic} emitter with its unparalleled angular resolution.

\section{Limitations} \label{sec:limitation}

In this section, we examine limitations of our model and their impact on our results.

\subsection{Acceleration Site} \label{subsec:acc}

We have focused on the case where $z\acc = z_s$, but in principle they could be different. If $z\acc \ll z_s$, we should take into account the particle cooling between $z\acc$ and $z_s$. In such a situation, the magnetic field at the acceleration site, $B\acc$, can also be different from the field at the emission region, $B_{\rm emit}$, which is derived by the spectral fitting. This difference would change our upper limits on $\eta\acc$ by a factor of $B_{\rm emit}/B\acc$ (see Eq.~\ref{eq:Emax_syn}). In particular, $B\acc$ could be smaller than $B_{\rm emit}$  otherwise we should see brighter synchrotron emission from the acceleration site. If this is the case, future observations should reveal fainter synchrotron emission from the acceleration site, placing better constraints on the magnetic field there. The difference between $B\acc$ and $B_{\rm emit}$ might be the reason why shocks are not yet resolved \citep{HAWC18}.

\subsection{Velocity in Knot Region} \label{subsec:v_z}

In our calculation, we have used $v_z = 0.26c$, which is determined at the jet base~\citep{Margon89,Eikenberry01}. The bulk velocity in the knot region is less certain from observations, but possibly be smaller than $0.26c$ because knots are located at large distances from the core~\citep{Goodall11a,Goodall11b,Monceau-Baroux14,Monceau-Baroux15,Panferov14,Panferov17,Bowler18}. The primarily effect of adopting a smaller bulk velocity would be flattening of the spectrum, because the transition from advection-dominated regime to fast cooling regime would occur at lower energy. This would produce a flat ($\Gamma_{\rm ph}\simeq2$) X-ray spectrum before a cutoff. In addition, the estimate on the size of the emitting regions would be proportionally changed for a different jet velocity. In other words, a better determination of both the spectrum and morphology in X-ray bands would be critical to constrain the bulk velocity in the knot region.

\section{Conclusions} \label{sec:conclusion}

Multiwavelength observations of the microquasar SS433 offer the potential for detailed studies on particle acceleration in astrophysical jets. In this paper, we first present a theoretical foundation to interpret nonthermal emission from astrophysical jets quantitatively. We then consider leptonic emission from the X-ray knots in SS433's jets. We use the same datasets as in \cite{HAWC18} and \cite{Xing19}, but treat the particle transport and evolution in the jet in more detail, and produce substantially different predictions. 

Our analysis produced three main results. First, leptonic models can explain the radio, X-ray and \ac{vhe} gamma-ray data for both the e1 and w1 regions. However, the GeV data remain unexplained for any reasonable parameter set, which indicates that they are mostly from different regions or mechanisms. Second, the efficiency of particle acceleration should be very high, $\eta\acc\lesssim10^2$, to explain the X-ray and TeV gamma-ray data. This could be realized by the diffusive shock acceleration, for a strong confinement case close to the Bohm limit, $\eta_g\sim1$. Such high efficiency of particle acceleration may imply that SS433 jets can also accelerate protons beyond a PeV. Third, future X-ray/MeV observations would be most critical to constrain models and better understand the acceleration processes. In particular, our models predict a spectral turnover and cutoff in this energy band.

We note that our models have broader implications that can be studied by future observations.

\begin{itemize}
\item  We have focused on the emission from e1 and w1 throughout this work. Our model can also predict emission from different regions by changing the parameter $z_s$ and $z_f$ in Eq.~(\ref{eq:dNdgamma}), provided that there is no effect of re-acceleration. As sketched in Fig.~\ref{fig:jet} and explained in Sec.~\ref{subsec:particle_qual} , in regions farther away from the binary, the synchrotron emission has spectral break steeper than expected from the cooling break. Interestingly, a hint of such a steep break is seen from observations of e2 and w2 regions \citep{SafiHarb97}. A better determination of the X-ray spectrum in these regions is the key to test this prediction.

\item We have focused on leptonic emission throughout this work. If protons are also accelerated in the jets, they may interact with the ambient medium to produce pionic gamma rays. Since the cooling time for protons, $\tau_{pp}$, is long, we may see emission from protons accumulated during the lifetime of SS433's jets, which should extend over much a larger region than X-ray knots. The jet kinetic power $\sim$~$10^{39}$~erg~s$^{-1}$ and system age $\sim$~$20$~kyr suggest that the jet has released the total energy of $E\jet$~$\sim$~$10^{51}$~erg. Assuming that 10$\%$ of this goes to nonthermal protons between 1~GeV and 1~PeV, and for an $E_p^{-2}$ injection spectrum, the proton energy would be $0.1E\jet$/ln($10^6$), yielding a TeV gamma-ray luminosity of $\sim$~$0.1E\jet /(3\tau_{pp}\ln(10^6))$. Thus, we could expect TeV gamma-ray flux of 
\be
F_\gamma \sim 10^{-13}(n_{\rm gas}/{\rm 0.2\ cm^{-3}})\ {\rm erg~s^{-1}~cm^{-2}}.
\ee
This could suggest that the hadronic gamma rays from the W50/SS433 system could also be detected at CTA and LHAASO. Furthermore, they may contribute to the VHE flux detected by HAWC, though it requires strong confinement of protons close to the emitting regions.

\end{itemize}

The first detection of SS433 in the \ac{vhe} regime has increased excitement in gamma-ray astronomy, by adding microquasars to a growing class of TeV sources. Our work highlights their importance as Galactic particle accelerators. Future observations with X-ray and \ac{vhe} gamma rays of SS433 and other microquasars should shed new light into our understanding of the high-energy sky.

\section*{Acknowledgements}
We are grateful to John F. Beacom, Valenti Bosch-Ramon, Ke Fang, Kazumi Kashiyama, Tim Linden, Chang Dong Rho, and Samar Safi-Harb for valuable comments on our manuscript. We also thank Felix Aharonian, Maxim Barkov, Shigeo S. Kimura and Kohta Murase for helpful discussions. We thank the anonymous referee for useful comments. This research made use of {\sc Matplotlib}~\citep{matplotlib}, {\sc Numpy}~\citep{numpybook,numpy} and {\sc Scipy}~\citep{scipy}. TS is supported by Research Fellowship of Japan Society for the Promotion of Science (JSPS), and also supported by JSPS KAKENHI Grant Number JP 18J20943. YI is supported by JSPS KAKENHI Grant Number JP16K13813, JP18H05458, JP19K14772, program of Leading Initiative for Excellent Young Researchers, MEXT, Japan, and RIKEN iTHEMS Program. DK is supported by JSPS KAKENHI Grant Numbers JP18H03722, JP24105007, and JP16H02170.

\twocolumngrid
\bibliography{ss433.bib}

\begin{thebibliography}{}
\expandafter\ifx\csname natexlab\endcsname\relax\def\natexlab#1{#1}\fi
\providecommand{\url}[1]{\href{#1}{#1}}

\bibitem[{{Abell} \& {Margon}(1979)}]{Abell79}
{Abell}, G.~O., \& {Margon}, B. 1979, \nat, 279, 701

\bibitem[{{Abeysekara} {et~al.}(2018){Abeysekara}, {Albert}, {Alfaro},
  {Alvarez}, {{\'A}lvarez}, {Arceo}, {Arteaga-Vel{\'a}zquez}, {Avila Rojas},
  {Ayala Solares}, {Belmont-Moreno}, {BenZvi}, {Brisbois}, {Caballero-Mora},
  {Capistr{\'a}n}, {Carrami{\~n}ana}, {Casanova}, {Castillo}, {Cotti},
  {Cotzomi}, {Couti{\~n}o de Le{\'o}n}, {De Le{\'o}n}, {De la Fuente},
  {D{\'\i}az-V{\'e}lez}, {Dichiara}, {Dingus}, {DuVernois}, {Ellsworth},
  {Engel}, {Espinoza}, {Fang}, {Fleischhack}, {Fraija}, {Galv{\'a}n-G{\'a}mez},
  {Garc{\'\i}a-Gonz{\'a}lez}, {Garfias}, {Gonz{\'a}lez Mu{\~n}oz},
  {Gonz{\'a}lez}, {Goodman}, {Hampel-Arias}, {Harding}, {Hernandez}, {Hinton},
  {Hona}, {Hueyotl-Zahuantitla}, {Hui}, {H{\"u}ntemeyer}, {Iriarte},
  {Jardin-Blicq}, {Joshi}, {Kaufmann}, {Kar}, {Kunde}, {Lauer}, {Lee},
  {Le{\'o}n Vargas}, {Li}, {Linnemann}, {Longinotti}, {Luis-Raya},
  {L{\'o}pez-Coto}, {Malone}, {Marinelli}, {Martinez}, {Martinez-Castellanos},
  {Mart{\'\i}nez-Castro}, {Matthews}, {Mirand a-Romagnoli}, {Moreno},
  {Mostaf{\'a}}, {Nayerhoda}, {Nellen}, {Newbold}, {Nisa}, {Noriega-Papaqui},
  {P{\'e}rez-P{\'e}rez}, {Pretz}, {Ren}, {Rho}, {Rivi{\`e}re},
  {Rosa-Gonz{\'a}lez}, {Rosenberg}, {Ruiz-Velasco}, {Salesa Greus}, {Sandoval},
  {Schneider}, {Schoorlemmer}, {Seglar Arroyo}, {Sinnis}, {Smith}, {Springer},
  {Surajbali}, {Taboada}, {Tibolla}, {Tollefson}, {Torres}, {Vianello},
  {Villase{\~n}or}, {Weisgarber}, {Werner}, {Westerhoff}, {Wood}, {Yapici},
  {Yodh}, {Zepeda}, {Zhang}, \& {Zhou}}]{HAWC18}
{Abeysekara}, A.~U., {Albert}, A., {Alfaro}, R., {et~al.} 2018, Nature, 562, 82

\bibitem[{{Acharya} {et~al.}(2019){Acharya}, {Agudo}, {Al Samarai}, {Alfaro},
  {Alfaro}, {Alispach}, {Alves Batista}, {Amans}, {Amato}, {Ambrosi},
  {Antolini}, {Antonelli}, {Aramo}, {Araya}, {Armstrong}, {Arqueros},
  {Arrabito}, {Asano}, {Ashley}, {Backes}, {Balazs}, {Balbo}, {Ballester},
  {Ballet}, {Bamba}, {Barkov}, {Barres de Almeida}, {Barrio}, {Bastieri},
  {Becherini}, {Belfiore}, {Benbow}, {Berge}, {Bernardini}, {Bernardini},
  {Bernardos}, {Bernl{\"o}hr}, {Bertucci}, {Biasuzzi}, {Bigongiari}, {Biland},
  {Bissaldi}, {Biteau}, {Blanch}, {Blazek}, {Boisson}, {Bolmont}, {Bonanno},
  {Bonardi}, {Bonavolont{\`a}}, {Bonnoli}, {Bosnjak}, {B{\"o}ttcher},
  {Braiding}, {Bregeon}, {Brill}, {Brown}, {Brun}, {Brunetti}, {Buanes},
  {Buckley}, {Bugaev}, {B{\"u}hler}, {Bulgarelli}, {Bulik}, {Burton},
  {Burtovoi}, {Busetto}, {Canestrari}, {Capalbi}, {Capitanio}, {Caproni},
  {Caraveo}, {C{\'a}rdenas}, {Carlile}, {Carosi}, {Carqu{\'\i}n}, {Carr},
  {Casanova}, {Cascone}, {Catalani}, {Catalano}, {Cauz}, {Cerruti}, {Chadwick},
  {Chaty}, {Chaves}, {Chen}, {Chen}, {Chernyakova}, {Chikawa}, {Christov},
  {Chudoba}, {Cie{\'s}lar}, {Coco}, {Colafrancesco}, {Colin}, {Conforti},
  {Connaughton}, {Conrad}, {Contreras}, {Cortina}, {Costa}, {Costantini},
  {Cotter}, {Covino}, {Crocker}, {Cuadra}, {Cuevas}, {Cumani}, {D'A{\`\i}},
  {D'Ammando}, {D'Avanzo}, {D'Urso}, {Daniel}, {Davids}, {Dawson}, {Dazzi}, {De
  Angelis}, {de C{\'a}ssia dos Anjos}, {De Cesare}, {De Franco}, {de Gouveia
  Dal Pino}, {de la Calle}, {de los Reyes Lopez}, {De Lotto}, {De Luca}, {De
  Lucia}, {de Naurois}, {de O{\~n}a Wilhelmi}, {De Palma}, {De Persio}, {de
  Souza}, {Deil}, {Del Santo}, {Delgado}, {della Volpe}, {Di Girolamo}, {Di
  Pierro}, {Di Venere}, {D{\'\i}az}, {Dib}, {Diebold}, {Djannati-Ata{\"\i}},
  {Dom{\'\i}nguez}, {Dominis Prester}, {Dorner}, {Doro}, {Drass}, {Dravins},
  {Dubus}, {Dwarkadas}, {Ebr}, {Eckner}, {Egberts}, {Einecke}, {Ekoume},
  {Els{\"a}sser}, {Ernenwein}, {Espinoza}, {Evoli}, {Fairbairn},
  {Falceta-Goncalves}, {Falcone}, {Farnier}, {Fasola}, {Fedorova}, {Fegan},
  {Fernand ez-Alonso}, {Fern{\'a}ndez-Barral}, {Ferrand}, {Fesquet},
  {Filipovic}, {Fioretti}, {Fontaine}, {Fornasa}, {Fortson}, {Freixas
  Coromina}, {Fruck}, {Fujita}, {Fukazawa}, {Funk}, {F{\"u}{\ss}ling},
  {Gabici}, {Gadola}, {Gallant}, {Garcia}, {Garcia L{\'o}pez}, {Garczarczyk},
  {Gaskins}, {Gasparetto}, {Gaug}, {Gerard}, {Giavitto}, {Giglietto}, {Giommi},
  {Giordano}, {Giro}, {Giroletti}, {Giuliani}, {Glicenstein}, {Gnatyk},
  {Godinovic}, {Goldoni}, {G{\'o}mez-Vargas}, {Gonz{\'a}lez}, {Gonz{\'a}lez},
  {G{\"o}tz}, {Graham}, {Grand i}, {Granot}, {Green}, {Greenshaw}, {Griffiths},
  {Gunji}, {Hadasch}, {Hara}, {Hardcastle}, {Hassan}, {Hayashi}, {Hayashida},
  {Heller}, {Helo}, {Hermann}, {Hinton}, {Hnatyk}, {Hofmann}, {Holder},
  {Horan}, {H{\"o}randel}, {Horns}, {Horvath}, {Hovatta}, {Hrabovsky},
  {Hrupec}, {Humensky}, {H{\"u}tten}, {Iarlori}, {Inada}, {Inome}, {Inoue},
  {Inoue}, {Inoue}, {Iocco}, {Ioka}, {Iori}, {Ishio}, {Iwamura}, {Jamrozy},
  {Janecek}, {Jankowsky}, {Jean}, {Jung-Richardt}, {Jurysek}, {Kaaret},
  {Karkar}, {Katagiri}, {Katz}, {Kawanaka}, {Kazanas}, {Kh{\'e}lifi}, {Kieda},
  {Kimeswenger}, {Kimura}, {Kisaka}, {Knapp}, {Kn{\"o}dlseder}, {Koch},
  {Kohri}, {Komin}, {Kosack}, {Kraus}, {Krause}, {Krau{\ss}}, {Kubo}, {Kukec
  Mezek}, {Kuroda}, {Kushida}, {La Palombara}, {Lamanna}, {Lang}, {Lapington},
  {Le Blanc}, {Leach}, {Lees}, {Lefaucheur}, {Leigui de Oliveira}, {Lenain},
  {Lico}, {Limon}, {Lindfors}, {Lohse}, {Lombardi}, {Longo}, {L{\'o}pez},
  {L{\'o}pez-Coto}, {Lu}, {Lucarelli}, {Luque-Escamilla}, {Lyard}, {Maccarone},
  {Maier}, {Majumdar}, {Malaguti}, {Mandat}, {Maneva}, {Manganaro}, {Mangano},
  {Marcowith}, {Mar{\'\i}n}, {Markoff}, {Mart{\'\i}}, {Martin},
  {Mart{\'\i}nez}, {Mart{\'\i}nez}, {Masetti}, {Masuda}, {Maurin}, {Maxted},
  {Mazin}, {Medina}, {Melandri}, {Mereghetti}, {Meyer}, {Minaya}, {Mirabal},
  {Mirzoyan}, {Mitchell}, {Mizuno}, {Moderski}, {Mohammed}, {Mohrmann},
  {Montaruli}, {Moralejo}, {Morcuende-Parrilla}, {Mori}, {Morlino}, {Morris},
  {Morselli}, {Moulin}, {Mukherjee}, {Mundell}, {Murach}, {Muraishi}, {Murase},
  {Nagai}, {Nagataki}, {Nagayoshi}, {Naito}, {Nakamori}, {Nakamura}, {Niemiec},
  {Nieto}, {Niko{\l}ajuk}, {Nishijima}, {Noda}, {Nosek}, {Novosyadlyj},
  {Nozaki}, {O'Brien}, {Oakes}, {Ohira}, {Ohishi}, {Ohm}, {Okazaki}, {Okumura},
  {Ong}, {Orienti}, {Orito}, {Osborne}, {Ostrowski}, {Otte}, {Oya}, {Padovani},
  {Paizis}, {Palatiello}, {Palatka}, {Paoletti}, {Paredes}, {Pareschi},
  {Parsons}, {Pe'er}, {Pech}, {Pedaletti}, {Perri}, {Persic}, {Petrashyk},
  {Petrucci}, {Petruk}, {Peyaud}, {Pfeifer}, {Piano}, {Pisarski}, {Pita},
  {Pohl}, {Polo}, {Pozo}, {Prandini}, {Prast}, {Principe}, {Prokhorov},
  {Prokoph}, {Prouza}, {P{\"u}hlhofer}, {Punch}, {P{\"u}rckhauer}, {Queiroz},
  {Quirrenbach}, {Rain{\`o}}, {Razzaque}, {Reimer}, {Reimer}, {Reisenegger},
  {Renaud}, {Rezaeian}, {Rhode}, {Ribeiro}, {Rib{\'o}}, {Richtler}, {Rico},
  {Rieger}, {Riquelme}, {Rivoire}, {Rizi}, {Rodriguez}, {Rodriguez Fernandez},
  {Rodr{\'\i}guez V{\'a}zquez}, {Rojas}, {Romano}, {Romeo}, {Rosado}, {Rovero},
  {Rowell}, {Rudak}, {Rugliancich}, {Rulten}, {Sadeh}, {Safi-Harb}, {Saito},
  {Sakaki}, {Sakurai}, {Salina}, {S{\'a}nchez-Conde}, {Sandaker}, {Sandoval},
  {Sangiorgi}, {Sanguillon}, {Sano}, {Santand er}, {Sarkar}, {Satalecka},
  {Saturni}, {Schioppa}, {Schlenstedt}, {Schneider}, {Schoorlemmer},
  {Schovanek}, {Schulz}, {Schussler}, {Schwanke}, {Sciacca}, {Scuderi},
  {Seitenzahl}, {Semikoz}, {Sergijenko}, {Servillat}, {Shalchi}, {Shellard},
  {Sidoli}, {Siejkowski}, {Sillanp{\"a}{\"a}}, {Sironi}, {Sitarek}, {Sliusar},
  {Slowikowska}, {Sol}, {Stamerra}, {Stani{\v{c}}}, {Starling}, {Stawarz},
  {Stefanik}, {Stephan}, {Stolarczyk}, {Stratta}, {Straumann}, {Suomijarvi},
  {Supanitsky}, {Tagliaferri}, {Tajima}, {Tavani}, {Tavecchio}, {Tavernet},
  {Tayabaly}, {Tejedor}, {Temnikov}, {Terada}, {Terrier}, {Terzic}, {Teshima},
  {Testa}, {Thoudam}, {Tian}, {Tibaldo}, {Tluczykont}, {Todero Peixoto},
  {Tokanai}, {Tomastik}, {Tonev}, {Tornikoski}, {Torres}, {Torresi}, {Tosti},
  {Tothill}, {Tovmassian}, {Travnicek}, {Trichard}, {Trifoglio}, {Troyano
  Pujadas}, {Tsujimoto}, {Umana}, {Vagelli}, {Vagnetti}, {Valentino},
  {Vallania}, {Valore}, {van Eldik}, {Vand enbroucke}, {Varner}, {Vasileiadis},
  {Vassiliev}, {V{\'a}zquez Acosta}, {Vecchi}, {Vega}, {Vercellone}, {Veres},
  {Vergani}, {Verzi}, {Vettolani}, {Viana}, {Vigorito}, {Villanueva}, {Voelk},
  {Vollhardt}, {Vorobiov}, {Vrastil}, {Vuillaume}, {Wagner}, {Wagner},
  {Walter}, {Ward}, {Warren}, {Watson}, {Werner}, {White}, {White},
  {Wierzcholska}, {Wilcox}, {Will}, {Williams}, {Wischnewski}, {Wood},
  {Yamamoto}, {Yamazaki}, {Yanagita}, {Yang}, {Yoshida}, {Yoshiike},
  {Yoshikoshi}, {Zacharias}, {Zaharijas}, {Zampieri}, {Zand anel}, {Zanin},
  {Zavrtanik}, {Zavrtanik}, {Zdziarski}, {Zech}, {Zechlin}, {Zhdanov},
  {Ziegler}, \& {Zorn}}]{CTAbook}
{Acharya}, B.~S., {Agudo}, I., {Al Samarai}, I., {et~al.} 2019, {Science with
  the Cherenkov Telescope Array}, doi:10.1142/10986

\bibitem[{{Aharonian} {et~al.}(1997){Aharonian}, {Atoyan}, \&
  {Kifune}}]{Aharonian97}
{Aharonian}, F.~A., {Atoyan}, A.~M., \& {Kifune}, T. 1997, \mnras, 291, 162

\bibitem[{{Aharonian} {et~al.}(2010){Aharonian}, {Kelner}, \&
  {Prosekin}}]{Aharonian10}
{Aharonian}, F.~A., {Kelner}, S.~R., \& {Prosekin}, A.~Y. 2010, \prd, 82,
  043002

\bibitem[{{Ahnen} {et~al.}(2018){Ahnen}, {Ansoldi}, {Antonelli}, {Arcaro},
  {Babi{\'c}}, {Banerjee}, {Bangale}, {Barres de Almeida}, {Barrio}, \&
  et~al.}]{MAGIC18}
{Ahnen}, M.~L., {Ansoldi}, S., {Antonelli}, L.~A., {et~al.} 2018, \aap, 612,
  A14

\bibitem[{{Amano} \& {Hoshino}(2012)}]{Amano12}
{Amano}, T., \& {Hoshino}, M. 2012, Astrophysics and Space Science Proceedings,
  33, 143

\bibitem[{{Antokhina} \& {Cherepashchuk}(1987)}]{Antokhina87}
{Antokhina}, E.~A., \& {Cherepashchuk}, A.~M. 1987, \sovast, 31, 295

\bibitem[{{Aramaki} {et~al.}(2020){Aramaki}, {Adrian}, {Karagiorgi}, \&
  {Odaka}}]{Aramaki2020}
{Aramaki}, T., {Adrian}, P. O.~H., {Karagiorgi}, G., \& {Odaka}, H. 2020,
  Astroparticle Physics, 114, 107

\bibitem[{{Araudo} {et~al.}(2015){Araudo}, {Bell}, \& {Blundell}}]{Araudo15}
{Araudo}, A.~T., {Bell}, A.~R., \& {Blundell}, K.~M. 2015, \apj, 806, 243

\bibitem[{{Atoyan} \& {Aharonian}(1999)}]{Atoyan99}
{Atoyan}, A.~M., \& {Aharonian}, F.~A. 1999, \mnras, 302, 253

\bibitem[{{Bai} {et~al.}(2019){Bai}, {Bi}, {Bi}, {Cao}, {Chen}, {Chen},
  {Chiavassa}, {Cui}, {Dai}, {della Volpe}, {Di Girolamo}, {Di Sciascio},
  {Fan}, {Giacalone}, {Guo}, {He}, {He}, {Heller}, {Huang}, {Huang}, {Jia},
  {Ksenofontov}, {Leahy}, {Li}, {Li}, {Liang}, {Lipari}, {Liu}, {Liu}, {Liu},
  {Ma}, {Martineau-Huynh}, {Martraire}, {Montaruli}, {Ruffolo}, {Stenkin},
  {Su}, {Tam}, {Tang}, {Tian}, {Vallania}, {Vernetto}, {Vigorito}, {Wang},
  {Wang}, {Wang}, {Wang}, {Wang}, {Wang}, {Wei}, {Wei}, {Wu}, {Wu}, {Wu},
  {Yan}, {Yang}, {Yang}, {Yao}, {Yin}, {Yuan}, {Zhang}, {Zhang}, {Zhang},
  {Zhang}, {Zhang}, {Zhang}, {Zhao}, {Zhou}, {Zhu}, \& {Zhu}}]{LHAASObook}
{Bai}, X., {Bi}, B.~Y., {Bi}, X.~J., {et~al.} 2019, arXiv e-prints,
  arXiv:1905.02773

\bibitem[{{Begelman} {et~al.}(2006){Begelman}, {King}, \&
  {Pringle}}]{Begelman06}
{Begelman}, M.~C., {King}, A.~R., \& {Pringle}, J.~E. 2006, \mnras, 370, 399

\bibitem[{{Bell}(2013)}]{Bell13}
{Bell}, A.~R. 2013, Astroparticle Physics, 43, 56

\bibitem[{{Bell} {et~al.}(2018){Bell}, {Araudo}, {Matthews}, \&
  {Blundell}}]{Bell18}
{Bell}, A.~R., {Araudo}, A.~T., {Matthews}, J.~H., \& {Blundell}, K.~M. 2018,
  \mnras, 473, 2364

\bibitem[{{Blumenthal} \& {Gould}(1970)}]{BG70}
{Blumenthal}, G.~R., \& {Gould}, R.~J. 1970, Reviews of Modern Physics, 42, 237

\bibitem[{{Blundell} \& {Bowler}(2004)}]{Blundell04}
{Blundell}, K.~M., \& {Bowler}, M.~G. 2004, \apj, 616, L159

\bibitem[{{Bordas} {et~al.}(2009){Bordas}, {Bosch-Ramon}, {Paredes}, \&
  {Perucho}}]{Bordas09}
{Bordas}, P., {Bosch-Ramon}, V., {Paredes}, J.~M., \& {Perucho}, M. 2009, \aap,
  497, 325

\bibitem[{{Bordas} {et~al.}(2017){Bordas}, {Sun}, {Yang}, {Kafexhiu}, \&
  {Aharonian}}]{Bordas17}
{Bordas}, P., {Sun}, X., {Yang}, R., {Kafexhiu}, E., \& {Aharonian}, F.~A.
  2017, in American Institute of Physics Conference Series, Vol. 1792, 6th
  International Symposium on High Energy Gamma-Ray Astronomy, 040020

\bibitem[{{Bordas} {et~al.}(2015){Bordas}, {Yang}, {Kafexhiu}, \&
  {Aharonian}}]{Bordas15}
{Bordas}, P., {Yang}, R., {Kafexhiu}, E., \& {Aharonian}, F. 2015, \apj, 807,
  L8

\bibitem[{{Bosch-Ramon} {et~al.}(2006){Bosch-Ramon}, {Romero}, \&
  {Paredes}}]{Bosch-Ramon06}
{Bosch-Ramon}, V., {Romero}, G.~E., \& {Paredes}, J.~M. 2006, \aap, 447, 263

\bibitem[{{Bowler} \& {Keppens}(2018)}]{Bowler18}
{Bowler}, M.~G., \& {Keppens}, R. 2018, \aap, 617, A29

\bibitem[{{Brinkmann} {et~al.}(1996){Brinkmann}, {Aschenbach}, \&
  {Kawai}}]{Brinkmann96}
{Brinkmann}, W., {Aschenbach}, B., \& {Kawai}, N. 1996, \aap, 312, 306

\bibitem[{{Brinkmann} \& {Kawai}(2000)}]{Brinkmann00}
{Brinkmann}, W., \& {Kawai}, N. 2000, \aap, 363, 640

\bibitem[{{Brinkmann} {et~al.}(2005){Brinkmann}, {Kotani}, \&
  {Kawai}}]{Brinkmann05}
{Brinkmann}, W., {Kotani}, T., \& {Kawai}, N. 2005, \aap, 431, 575

\bibitem[{{Brinkmann} {et~al.}(2007){Brinkmann}, {Pratt}, {Rohr}, {Kawai}, \&
  {Burwitz}}]{Brinkmann07}
{Brinkmann}, W., {Pratt}, G.~W., {Rohr}, S., {Kawai}, N., \& {Burwitz}, V.
  2007, \aap, 463, 611

\bibitem[{{Broderick} {et~al.}(2018){Broderick}, {Fender}, {Miller-Jones},
  {Trushkin}, {Stewart}, {Anderson}, {Staley}, {Blundell}, {Pietka}, {Markoff},
  {Rowlinson}, {Swinbank}, {van der Horst}, {Bell}, {Breton}, {Carbone},
  {Corbel}, {Eisl{\"o}ffel}, {Falcke}, {Grie{\ss}meier}, {Hessels},
  {Kondratiev}, {Law}, {Molenaar}, {Serylak}, {Stappers}, {van Leeuwen},
  {Wijers}, {Wijnands}, {Wise}, \& {Zarka}}]{LOFAR18}
{Broderick}, J.~W., {Fender}, R.~P., {Miller-Jones}, J.~C.~A., {et~al.} 2018,
  \mnras, 475, 5360

\bibitem[{{Bykov} {et~al.}(2012){Bykov}, {Pavlov}, {Artemyev}, \&
  {Uvarov}}]{Bykov12}
{Bykov}, A.~M., {Pavlov}, G.~G., {Artemyev}, A.~V., \& {Uvarov}, Y.~A. 2012,
  \mnras, 421, L67

\bibitem[{{Cherepashchuk} {et~al.}(2019){Cherepashchuk}, {Postnov}, \&
  {Belinski}}]{Cherepashchuk19}
{Cherepashchuk}, A.~M., {Postnov}, K.~A., \& {Belinski}, A.~A. 2019, \mnras,
  485, 2638

\bibitem[{{Cherepashchuk} {et~al.}(2013){Cherepashchuk}, {Sunyaev}, {Molkov},
  {Antokhina}, {Postnov}, \& {Bogomazov}}]{Cherepashchuk13}
{Cherepashchuk}, A.~M., {Sunyaev}, R.~A., {Molkov}, S.~V., {et~al.} 2013,
  \mnras, 436, 2004

\bibitem[{{De Angelis} {et~al.}(2017){De Angelis}, {Tatischeff}, {Tavani},
  {Oberlack}, {Grenier}, {Hanlon}, {Walter}, {Argan}, {von Ballmoos},
  {Bulgarelli}, {Donnarumma}, {Hernanz}, {Kuvvetli}, {Pearce}, {Zdziarski},
  {Aboudan}, {Ajello}, {Ambrosi}, {Bernard}, {Bernardini}, {Bonvicini},
  {Brogna}, {Branchesi}, {Budtz-Jorgensen}, {Bykov}, {Campana}, {Cardillo},
  {Coppi}, {De Martino}, {Diehl}, {Doro}, {Fioretti}, {Funk}, {Ghisellini},
  {Grove}, {Hamadache}, {Hartmann}, {Hayashida}, {Isern}, {Kanbach}, {Kiener},
  {Kn{\"o}dlseder}, {Labanti}, {Laurent}, {Limousin}, {Longo}, {Mannheim},
  {Marisaldi}, {Martinez}, {Mazziotta}, {McEnery}, {Mereghetti}, {Minervini},
  {Moiseev}, {Morselli}, {Nakazawa}, {Orleanski}, {Paredes}, {Patricelli},
  {Peyr{\'e}}, {Piano}, {Pohl}, {Ramarijaona}, {Rando}, {Reichardt},
  {Roncadelli}, {Silva}, {Tavecchio}, {Thompson}, {Turolla}, {Ulyanov},
  {Vacchi}, {Wu}, \& {Zoglauer}}]{DeAngelis2017}
{De Angelis}, A., {Tatischeff}, V., {Tavani}, M., {et~al.} 2017, Experimental
  Astronomy, 44, 25

\bibitem[{{Derishev} \& {Aharonian}(2019)}]{2019arXiv190711663D}
{Derishev}, E., \& {Aharonian}, F. 2019, arXiv e-prints, arXiv:1907.11663

\bibitem[{{Dermer} \& {Menon}(2009)}]{Dermer09}
{Dermer}, C.~D., \& {Menon}, G. 2009, {High Energy Radiation from Black Holes:
  Gamma Rays, Cosmic Rays, and Neutrinos}

\bibitem[{{Downes} {et~al.}(1986){Downes}, {Pauls}, \& {Salter}}]{Downes86}
{Downes}, A.~J.~B., {Pauls}, T., \& {Salter}, C.~J. 1986, \mnras, 218, 393

\bibitem[{{Dubner} {et~al.}(1998){Dubner}, {Holdaway}, {Goss}, \&
  {Mirabel}}]{Dubner98}
{Dubner}, G.~M., {Holdaway}, M., {Goss}, W.~M., \& {Mirabel}, I.~F. 1998, \aj,
  116, 1842

\bibitem[{{Eikenberry} {et~al.}(2001){Eikenberry}, {Cameron}, {Fierce}, {Kull},
  {Dror}, {Houck}, \& {Margon}}]{Eikenberry01}
{Eikenberry}, S.~S., {Cameron}, P.~B., {Fierce}, B.~W., {et~al.} 2001, \apj,
  561, 1027

\bibitem[{{Fabian} \& {Rees}(1979)}]{Fabian79}
{Fabian}, A.~C., \& {Rees}, M.~J. 1979, \mnras, 187, 13P

\bibitem[{{Fabrika}(2004)}]{Fabrika04}
{Fabrika}, S. 2004, Astrophysics and Space Physics Reviews, 12, 1

\bibitem[{{Geldzahler} {et~al.}(1980){Geldzahler}, {Pauls}, \&
  {Salter}}]{Geldzahler80}
{Geldzahler}, B.~J., {Pauls}, T., \& {Salter}, C.~J. 1980, \aap, 84, 237

\bibitem[{{Goodall} {et~al.}(2011{\natexlab{a}}){Goodall}, {Alouani-Bibi}, \&
  {Blundell}}]{Goodall11a}
{Goodall}, P.~T., {Alouani-Bibi}, F., \& {Blundell}, K.~M. 2011{\natexlab{a}},
  \mnras, 414, 2838

\bibitem[{{Goodall} {et~al.}(2011{\natexlab{b}}){Goodall}, {Blundell}, \& {Bell
  Burnell}}]{Goodall11b}
{Goodall}, P.~T., {Blundell}, K.~M., \& {Bell Burnell}, S.~J.
  2011{\natexlab{b}}, \mnras, 414, 2828

\bibitem[{{Green}(2004)}]{Green04}
{Green}, D.~A. 2004, Bulletin of the Astronomical Society of India, 32, 335

\bibitem[{{Gupta} {et~al.}(2006){Gupta}, {B{\"o}ttcher}, \& {Dermer}}]{Gupta06}
{Gupta}, S., {B{\"o}ttcher}, M., \& {Dermer}, C.~D. 2006, \apj, 644, 409

\bibitem[{{Heinz} \& {Sunyaev}(2002)}]{Heinz02}
{Heinz}, S., \& {Sunyaev}, R. 2002, \aap, 390, 751

\bibitem[{{Hillas}(1984)}]{1984ARA&A..22..425H}
{Hillas}, A.~M. 1984, \araa, 22, 425

\bibitem[{{Hillwig} \& {Gies}(2008)}]{Hillwig08}
{Hillwig}, T.~C., \& {Gies}, D.~R. 2008, \apjl, 676, L37

\bibitem[{{Hillwig} {et~al.}(2004){Hillwig}, {Gies}, {Huang}, {McSwain},
  {Stark}, {van der Meer}, \& {Kaper}}]{Hillwig04}
{Hillwig}, T.~C., {Gies}, D.~R., {Huang}, W., {et~al.} 2004, \apj, 615, 422

\bibitem[{Hunter(2007)}]{matplotlib}
Hunter, J.~D. 2007, Computing in Science \& Engineering, 9, 90

\bibitem[{{Inoue} \& {Tanaka}(2016)}]{Inoue2016}
{Inoue}, Y., \& {Tanaka}, Y.~T. 2016, \apj, 828, 13

\bibitem[{Jones {et~al.}(2001--)Jones, Oliphant, Peterson, {et~al.}}]{scipy}
Jones, E., Oliphant, T., Peterson, P., {et~al.} 2001--, {SciPy}: Open source
  scientific tools for {Python}".
\newblock \url{http://www.scipy.org/}

\bibitem[{{Kar}(2017)}]{Veritas17}
{Kar}, P. 2017, International Cosmic Ray Conference, 35, 713

\bibitem[{{Kelner} {et~al.}(2013){Kelner}, {Aharonian}, \&
  {Khangulyan}}]{Kelner13}
{Kelner}, S.~R., {Aharonian}, F.~A., \& {Khangulyan}, D. 2013, \apj, 774, 61

\bibitem[{{Khangulyan} {et~al.}(2014){Khangulyan}, {Aharonian}, \&
  {Kelner}}]{Khangulyan14}
{Khangulyan}, D., {Aharonian}, F.~A., \& {Kelner}, S.~R. 2014, \apj, 783, 100

\bibitem[{{Khangulyan} {et~al.}(2018){Khangulyan}, {Bosch-Ramon}, \&
  {Uchiyama}}]{2018MNRAS.481.1455K}
{Khangulyan}, D., {Bosch-Ramon}, V., \& {Uchiyama}, Y. 2018, \mnras, 481, 1455

\bibitem[{{Kotani} {et~al.}(1996){Kotani}, {Kawai}, {Matsuoka}, \&
  {Brinkmann}}]{Kotani96}
{Kotani}, T., {Kawai}, N., {Matsuoka}, M., \& {Brinkmann}, W. 1996, \pasj, 48,
  619

\bibitem[{{Kubota} {et~al.}(2010){Kubota}, {Ueda}, {Fabrika}, {Medvedev},
  {Barsukova}, {Sholukhova}, \& {Goranskij}}]{Kubota10}
{Kubota}, K., {Ueda}, Y., {Fabrika}, S., {et~al.} 2010, \apj, 709, 1374

\bibitem[{{Lockman} {et~al.}(2007){Lockman}, {Blundell}, \& {Goss}}]{Lockman07}
{Lockman}, F.~J., {Blundell}, K.~M., \& {Goss}, W.~M. 2007, \mnras, 381, 881

\bibitem[{{Margon} \& {Anderson}(1989)}]{Margon89}
{Margon}, B., \& {Anderson}, S.~F. 1989, \apj, 347, 448

\bibitem[{{Marshall} {et~al.}(2002){Marshall}, {Canizares}, \&
  {Schulz}}]{Marshall02}
{Marshall}, H.~L., {Canizares}, C.~R., \& {Schulz}, N.~S. 2002, \apj, 564, 941

\bibitem[{{McEnery} {et~al.}(2019){McEnery}, {Abel Barrio}, {Agudo}, {Ajello},
  {{\'A}lvarez}, {Ansoldi}, {Anton}, {Auricchio}, {Stephen}, {Baldini},
  {Bambi}, {Baring}, {Barres}, {Bastieri}, {Beacom}, {Beckmann}, {Bednarek},
  {Bernard}, {Bissaldi}, {Bloser}, {Blumer}, {Boettcher}, {Boggs},
  {Bolotnikov}, {Bottacini}, {Bozhilov}, {Bozzo}, {Briggs}, {Buckley}, {Buson},
  {Campana}, {Caputo}, {Cardillo}, {Caroli}, {Castro}, {Cenko}, {Charles},
  {Chen}, {Cheung}, {Ciprini}, {Coppi}, {Curado da Silva}, {Cutini}, {D'Ammand
  o}, {De Angelis}, {De Becker}, {De Nolfo}, {Del Sordo}, {Di Mauro}, {Di
  Venere}, {Dietrich}, {Digel}, {Dominguez}, {Doro}, {Ferrara}, {Fields},
  {Finke}, {Foffano}, {Fryer}, {Fukazawa}, {Funk}, {Gasparrini}, {Gelfand},
  {Georganopoulos}, {Giordano}, {Giuliani}, {Gouiffes}, {Grefenstette},
  {Grenier}, {Griffin}, {Grove}, {Guiriec}, {Harding}, {Harding}, {Hartmann},
  {Hays}, {Hernanz}, {Hewitt}, {Holder}, {Hui}, {Inglis}, {Johnson}, {Jones},
  {Kanbach}, {Kargaltsev}, {Kaufmann}, {Kerr}, {Kierans}, {Kislat}, {Klimenko},
  {Knodlseder}, {Kocveski}, {Kopp}, {Krawczynsiki}, {Krizmanic}, {Kubo},
  {Kurahashi Neilson}, {Laurent}, {Lenain}, {Li}, {Lien}, {Linden}, {Lommler},
  {Longo}, {Lovellette}, {L{\'o}pez}, {Manousakis}, {Marcotulli}, {Marcowith},
  {Martinez}, {McConnell}, {Metcalfe}, {Meyer}, {Meyer}, {Mignani}, {Mitchell},
  {Mizuno}, {Moiseev}, {Morcuende}, {Moskalenko}, {Moss}, {Nakazawa},
  {Mazziotta}, {Oberlack}, {Ohno}, {Oikonomou}, {Ojha}, {Omodei}, {Orlando},
  {Otte}, {Paliya}, {Parker}, {Patricelli}, {Perkins}, {Petropoulou},
  {Pittori}, {Pohl}, {Porter}, {Prandini}, {Prescod-Weinstein}, {Racusin},
  {Rand o}, {Rani}, {Rib{\'o}}, {Rodi}, {Sanchez-Conde}, {Saz Parkinson},
  {Schirato}, {Shawhan}, {Shrader}, {Smith}, {Smith}, {Stamerra}, {Stawarz},
  {Strong}, {Stumke}, {Tajima}, {Takahashi}, {Tanaka}, {Tatischeff}, {The},
  {Thompson}, {Tibaldo}, {Tomsick}, {Uhm}, {Venters}, {Vestrand}, {Vianello},
  {Wadiasingh}, {Walter}, {Wang}, {Williams}, {Wilson-Hodge}, {Wood}, {Woolf},
  {Wulf}, {Younes}, {Zampieri}, {Zane}, {Zhang}, {Zhang}, {Zimmer}, {Zoglauer},
  \& {van der Horst}}]{AMEGO2019}
{McEnery}, J., {Abel Barrio}, J., {Agudo}, I., {et~al.} 2019, arXiv e-prints,
  arXiv:1907.07558

\bibitem[{{Medvedev} \& {Fabrika}(2010)}]{Medvedev10}
{Medvedev}, A., \& {Fabrika}, S. 2010, \mnras, 402, 479

\bibitem[{{Molina} \& {Bosch-Ramon}(2018)}]{Molina18}
{Molina}, E., \& {Bosch-Ramon}, V. 2018, Astronomy and Astrophysics, 618, A146

\bibitem[{{Monceau-Baroux} {et~al.}(2014){Monceau-Baroux}, {Porth}, {Meliani},
  \& {Keppens}}]{Monceau-Baroux14}
{Monceau-Baroux}, R., {Porth}, O., {Meliani}, Z., \& {Keppens}, R. 2014, \aap,
  561, A30

\bibitem[{{Monceau-Baroux} {et~al.}(2015){Monceau-Baroux}, {Porth}, {Meliani},
  \& {Keppens}}]{Monceau-Baroux15}
---. 2015, \aap, 574, A143

\bibitem[{Oliphant(2006--)}]{numpybook}
Oliphant, T. 2006--, {NumPy}: A guide to {NumPy}.
\newblock \url{http://www.numpy.org/}

\bibitem[{{Orellana} {et~al.}(2007){Orellana}, {Bordas}, {Bosch-Ramon},
  {Romero}, \& {Paredes}}]{Orellana07}
{Orellana}, M., {Bordas}, P., {Bosch-Ramon}, V., {Romero}, G.~E., \& {Paredes},
  J.~M. 2007, \aap, 476, 9

\bibitem[{{O'Sullivan} {et~al.}(2009){O'Sullivan}, {Reville}, \&
  {Taylor}}]{OSullivan09}
{O'Sullivan}, S., {Reville}, B., \& {Taylor}, A.~M. 2009, \mnras, 400, 248

\bibitem[{{Panferov}(2014)}]{Panferov14}
{Panferov}, A. 2014, \aap, 562, A130

\bibitem[{{Panferov}(2017)}]{Panferov17}
{Panferov}, A.~A. 2017, \aap, 599, A77

\bibitem[{{Pepe} {et~al.}(2015){Pepe}, {Vila}, \& {Romero}}]{Pepe15}
{Pepe}, C., {Vila}, G.~S., \& {Romero}, G.~E. 2015, \aap, 584, A95

\bibitem[{{Perucho} \& {Bosch-Ramon}(2008)}]{Perucho08}
{Perucho}, M., \& {Bosch-Ramon}, V. 2008, \aap, 482, 917

\bibitem[{{Popescu} {et~al.}(2017){Popescu}, {Yang}, {Tuffs}, {Natale},
  {Rushton}, \& {Aharonian}}]{Popescu17}
{Popescu}, C.~C., {Yang}, R., {Tuffs}, R.~J., {et~al.} 2017, \mnras, 470, 2539

\bibitem[{{Porter} {et~al.}(2017){Porter}, {J{\'o}hannesson}, \&
  {Moskalenko}}]{Porter17}
{Porter}, T.~A., {J{\'o}hannesson}, G., \& {Moskalenko}, I.~V. 2017, \apj, 846,
  67

\bibitem[{{Rasul} {et~al.}(2019){Rasul}, {Chadwick}, {Graham}, \&
  {Brown}}]{Rasul19}
{Rasul}, K., {Chadwick}, P.~M., {Graham}, J.~A., \& {Brown}, A.~M. 2019,
  \mnras, 485, 2970

\bibitem[{{Reynoso} \& {Carulli}(2019)}]{Reynoso19}
{Reynoso}, M.~M., \& {Carulli}, A.~M. 2019, Astroparticle Physics, 109, 25

\bibitem[{{Reynoso} {et~al.}(2008){Reynoso}, {Romero}, \&
  {Christiansen}}]{Reynoso08}
{Reynoso}, M.~M., {Romero}, G.~E., \& {Christiansen}, H.~R. 2008, \mnras, 387,
  1745

\bibitem[{{Romero} {et~al.}(2003){Romero}, {Torres}, {Kaufman Bernad{\'o}}, \&
  {Mirabel}}]{Romero03}
{Romero}, G.~E., {Torres}, D.~F., {Kaufman Bernad{\'o}}, M.~M., \& {Mirabel},
  I.~F. 2003, \aap, 410, L1

\bibitem[{{Romero} \& {Vila}(2008)}]{Romero08}
{Romero}, G.~E., \& {Vila}, G.~S. 2008, \aap, 485, 623

\bibitem[{{Safi-Harb} \& {{\"O}gelman}(1997)}]{SafiHarb97}
{Safi-Harb}, S., \& {{\"O}gelman}, H. 1997, \apj, 483, 868

\bibitem[{{Safi-Harb} \& {Petre}(1999)}]{SafiHarb99}
{Safi-Harb}, S., \& {Petre}, R. 1999, \apj, 512, 784

\bibitem[{{Stage} {et~al.}(2006){Stage}, {Allen}, {Houck}, \&
  {Davis}}]{Stage06}
{Stage}, M.~D., {Allen}, G.~E., {Houck}, J.~C., \& {Davis}, J.~E. 2006, Nature
  Physics, 2, 614

\bibitem[{{Sun} {et~al.}(2019){Sun}, {Yang}, {Liu}, {Xi}, \& {Wang}}]{Sun19}
{Sun}, X.-N., {Yang}, R.-Z., {Liu}, B., {Xi}, S.-Q., \& {Wang}, X.-Y. 2019,
  \aap, 626, A113

\bibitem[{{Tanada} {et~al.}(2019){Tanada}, {Kataoka}, \& {Inoue}}]{Tanada2019}
{Tanada}, K., {Kataoka}, J., \& {Inoue}, Y. 2019, \apj, 878, 139

\bibitem[{{Tsuji} {et~al.}(2019){Tsuji}, {Uchiyama}, {Aharonian}, {Berge},
  {Higurashi}, {Krivonos}, \& {Tanaka}}]{Tsuji19}
{Tsuji}, N., {Uchiyama}, Y., {Aharonian}, F., {et~al.} 2019, \apj, 877, 96

\bibitem[{{Uchiyama} {et~al.}(2007){Uchiyama}, {Aharonian}, {Tanaka},
  {Takahashi}, \& {Maeda}}]{Uchiyama2007}
{Uchiyama}, Y., {Aharonian}, F.~A., {Tanaka}, T., {Takahashi}, T., \& {Maeda},
  Y. 2007, \nat, 449, 576

\bibitem[{{Vaidya} {et~al.}(2018){Vaidya}, {Mignone}, {Bodo}, {Rossi}, \&
  {Massaglia}}]{2018ApJ...865..144V}
{Vaidya}, B., {Mignone}, A., {Bodo}, G., {Rossi}, P., \& {Massaglia}, S. 2018,
  \apj, 865, 144

\bibitem[{{van der Walt} {et~al.}(2011){van der Walt}, {Colbert}, \&
  {Varoquaux}}]{numpy}
{van der Walt}, S., {Colbert}, S.~C., \& {Varoquaux}, G. 2011, Computing in
  Science Engineering, 13, 22

\bibitem[{{Vila} \& {Romero}(2010)}]{Vila10}
{Vila}, G.~S., \& {Romero}, G.~E. 2010, \mnras, 403, 1457

\bibitem[{{Waisberg} {et~al.}(2019){Waisberg}, {Dexter}, {Olivier-Petrucci},
  {Dubus}, \& {Perraut}}]{Waisberg19}
{Waisberg}, I., {Dexter}, J., {Olivier-Petrucci}, P., {Dubus}, G., \&
  {Perraut}, K. 2019, \aap, 624, A127

\bibitem[{{Watson} {et~al.}(1983){Watson}, {Willingale}, {Grindlay}, \&
  {Seward}}]{Watson83}
{Watson}, M.~G., {Willingale}, R., {Grindlay}, J.~E., \& {Seward}, F.~D. 1983,
  \apj, 273, 688

\bibitem[{{Webb}(1989)}]{1989ApJ...340.1112W}
{Webb}, G.~M. 1989, \apj, 340, 1112

\bibitem[{{Xing} {et~al.}(2019){Xing}, {Wang}, {Zhang}, {Chen}, \&
  {Jithesh}}]{Xing19}
{Xing}, Y., {Wang}, Z., {Zhang}, X., {Chen}, Y., \& {Jithesh}, V. 2019, \apj,
  872, 25

\bibitem[{{Yamauchi} {et~al.}(1994){Yamauchi}, {Kawai}, \& {Aoki}}]{Yamauchi94}
{Yamauchi}, S., {Kawai}, N., \& {Aoki}, T. 1994, \pasj, 46, L109

\bibitem[{{Zdziarski} {et~al.}(2014){Zdziarski}, {Pjanka}, {Sikora}, \&
  {Stawarz}}]{Zdziarski14}
{Zdziarski}, A.~A., {Pjanka}, P., {Sikora}, M., \& {Stawarz}, {\L}. 2014,
  \mnras, 442, 3243

\bibitem[{{Zealey} {et~al.}(1980){Zealey}, {Dopita}, \& {Malin}}]{Zealey80}
{Zealey}, W.~J., {Dopita}, M.~A., \& {Malin}, D.~F. 1980, \mnras, 192, 731

\bibitem[{{Zhang} {et~al.}(2018){Zhang}, {Li}, {Xiang}, \& {Lu}}]{Zhang18}
{Zhang}, J.-F., {Li}, Z.-R., {Xiang}, F.-Y., \& {Lu}, J.-F. 2018, \mnras, 473,
  3211

\end{thebibliography}

\end{document}